\documentclass[aastex,superscriptaddress,showpacs,letterpaper,showkeys,preprintnumbers,altaffilletter,amssymb,amsmath,amsfonts,aps]{emulateapj}

\usepackage[utf8]{inputenc}
\usepackage[T1]{fontenc}
\usepackage{ae,aecompl} 
\usepackage{graphicx}
\usepackage{amsmath}
\usepackage{color}
\usepackage{amssymb}
\usepackage{latexsym}
\usepackage{wasysym}
\usepackage{psfrag}
\usepackage{comment}
\usepackage{ifthen}
\usepackage[colorlinks=true]{hyperref}

\usepackage{natbib}
\def\be{\begin{equation}}
\def\ee{\end{equation}}
\def\ben{$$}
\def\een{$$}
\def\bea{\begin{eqnarray}}
\def\eea{\end{eqnarray}}
\def\bean{\begin{eqnarray*}}
\def\eean{\end{eqnarray*}}

\def\bi{\begin{itemize}}
\def\ei{\end{itemize}}
\def\ben{\begin{enumerate}}
\def\een{\end{enumerate}}

\def\xp{{\textsc{X-Pipeline}}}

\newcommand\unit[2]{\mbox{#1\,#2}}

\newcommand{\lsem}{[\![}
\newcommand{\rsem}{]\!]}
\newcommand{\makevisible}[1]{\textbf{#1}}
\newcommand{\switch}[1]{
  \ifthenelse{\equal{#1}{0}}{\renewcommand{\makevisible}[1]{}}{}}
\switch{0} 

\newcommand{\distNSNS}{16}
\newcommand{\distNSBH}{28}
\newcommand{\nTotGRB}{404}
\newcommand{\nBurstGRB}{150}
\newcommand{\nAnalyzedGRB}{154}

\newcommand{\dccversion}{10}

\shorttitle{Search for GWs associated with GRBs using LIGO and Virgo}
\shortauthors{Abadie et al.}

\begin{document}
\title{Search for gravitational waves associated with
  gamma-ray bursts during LIGO science run 6 and Virgo science runs 2
  and 3}
\author{J.~Abadie$^{1}$, 
B.~P.~Abbott$^{1}$, 
R.~Abbott$^{1}$, 
T.~D.~Abbott$^{2}$, 
M.~Abernathy$^{3}$, 
T.~Accadia$^{4}$, 
F.~Acernese$^{5ac}$, 
C.~Adams$^{6}$, 
R.~X.~Adhikari$^{1}$, 
C.~Affeldt$^{7,8}$
% ...
}
\address{$^{1}$LIGO - California Institute of Technology, Pasadena, CA  91125, USA }
\address{$^{2}$California State University Fullerton, Fullerton CA 92831 USA}
\address{$^{3}$SUPA, University of Glasgow, Glasgow, G12 8QQ, United Kingdom }
\address{$^{4}$Laboratoire d'Annecy-le-Vieux de Physique des Particules (LAPP), Universit\'e de Savoie, CNRS/IN2P3, F-74941 Annecy-Le-Vieux, France}
\address{$^{5}$INFN, Sezione di Napoli $^a$; Universit\`a di Napoli 'Federico II'$^b$ Complesso Universitario di Monte S.Angelo, I-80126 Napoli; Universit\`a di Salerno, Fisciano, I-84084 Salerno$^c$, Italy}
\address{$^{6}$LIGO - Livingston Observatory, Livingston, LA  70754, USA }
\address{$^{7}$Albert-Einstein-Institut, Max-Planck-Institut f\"ur Gravitationsphysik, D-30167 Hannover, Germany}
\address{$^{8}$Leibniz Universit\"at Hannover, D-30167 Hannover, Germany }
% ...
\affil{(The LIGO Scientific Collaboration and the Virgo Collaboration)}

\author{
M.~S.~Briggs$^{92}$,
V.~Connaughton$^{92}$,
K.~C.~Hurley$^{93}$,
P.~A.~Jenke$^{94}$,
A.~{von Kienlin}$^{95}$,
A.~Rau$^{95}$,
X.-L.~Zhang$^{95}$
}
\address{$^{92}$CSPAR, University of Alabama in Huntsville, Huntsville, Alabama, USA}
\address{$^{93}$University of California-Berkeley, Space Sciences Lab, 7 Gauss Way, Berkeley, CA 94720, USA}
\address{$^{94}$Marshall Space Flight Center  Huntsville, AL 35811, United States}
\address{$^{95}$Max-Planck-Institut f\"ur extraterrestrische Physik, Giessenbachstra{\ss}e, 85748 Garching, Germany}

\begin{abstract}

We present the results of a search for gravitational waves associated
with \nAnalyzedGRB\
gamma-ray bursts (GRBs) that were detected by satellite-based
gamma-ray experiments in 2009-2010, during the sixth LIGO science run and the second
and third Virgo science runs. 
We perform two distinct searches: a modeled search for coalescences of
either two neutron stars or a neutron star and black hole; and a search for
generic, unmodeled gravitational-wave bursts. 
We find no evidence for
gravitational-wave counterparts, either with any individual GRB in this sample 
or with the population as a whole. 
For all GRBs we place lower bounds on the distance to the progenitor,
under the optimistic assumption of a gravitational-wave emission energy of
$10^{-2}\,\mathrm{M_\odot c^2}$ at $150\,\mathrm{Hz}$, with a median
limit of 17\,Mpc.  For short hard GRBs 
we place exclusion distances on binary neutron star and neutron
star--black hole progenitors, using astrophysically motivated priors on 
the source parameters,
with median values of \distNSNS\,Mpc and \distNSBH\,Mpc respectively.
These distance limits, while significantly larger than for a search that is
not aided by GRB satellite observations, are not large enough to expect 
a coincidence with a GRB. 
However, projecting these exclusions to the sensitivities of
Advanced LIGO and Virgo, which should begin operation in 2015,
we find that the detection of gravitational waves associated with GRBs
will become quite possible.

\end{abstract}



\section{Introduction} 
\label{sec:intro} 

Gamma-ray bursts (GRBs) are intense flashes of $\gamma$-rays which 
are observed approximately once per day and are isotropically distributed 
over the sky \citep[see, e.g.][and references therein]{Meszaros:2006rc}.
The variability of the bursts on time scales as short as a 
millisecond indicates that the sources are very compact, while the 
identification of host galaxies and the measurement of redshifts for 
more than 200 bursts have shown that GRBs are of extra-galactic origin.  

GRBs are grouped into two broad classes by their characteristic duration and
spectral hardness \citep{ck93}.  Long GRBs ($\gtrsim$ 2 s, with
softer spectra), are related to the collapse of massive stars
with highly rotating cores
\citep[see e.g. reviews][]{2011AN....332..434M,2011arXiv1104.2274H}.
The extreme
core-collapse scenarios leading to GRBs result in the formation of a
stellar-mass black hole with an accretion disk 
or of a highly-magnetized neutron star; for a review see
\citet{Woosley11} and references therein.
In both cases the
emission of gravitational waves (GWs) is expected, though the
amount of emission is highly uncertain.  

The progenitors of most short GRBs ($\lesssim$ 2 s, with harder spectra) are
widely thought to be mergers of neutron star-neutron star or neutron star-black
hole binaries
\citep[see, e.g.][]{schramm89,Narayan:1992iy,nakar-2007,Gehrels09}, 
though up to a few percent 
may be due to giant flares from a local distribution of soft-gamma repeaters
\citep{duncan92,2005Natur.438..991T,palmer05,NaGaFo:06,Frederiks07,mazets07,2009MNRAS.395.1515C,2010MNRAS.403..342H}.
The mergers, referred to here as 
compact binary coalescences, are expected to be strong GW radiators \citep{thorne.k:1987}.  The 
detection of gravitational waves associated with a short GRB would
provide direct evidence that the progenitor is indeed a compact binary.
With such a detection it would be possible to measure component masses
\citep{Finn:1992xs,Cutler:1994ys} and spins
\citep{Poisson:1995ef}, constrain neutron star equations of state
\citep{2000PhRvL..84.3519V,Flanagan:2007ix,2010PhRvD..81l3016H,Read:2009yp,2011arXiv1109.3402L,2011PhRvD..84j4017P}, test general relativity in the
strong-field regime \citep{Will:2005va}, and measure calibration-free
luminosity distances \citep{Schutz:1986gp,1993ApJ...411L...5C,Dragoljub93,Dalal:2006qt,Nissanke:2009kt}, which allow the measurement of
the Hubble expansion and dark energy.

Several searches for gravitational waves associated with gamma-ray bursts 
have been performed using data from LIGO and Virgo 
\citep{abbottgrb05,burstGrbS234,Ac_etal:07,Ac_etal:08}.  Most recently, data 
from the fifth LIGO science run and the first Virgo science run  
were analyzed to search for coalescence signals or unmodeled gravitational-wave 
bursts associated with 137 GRBs from 2005-2007 \citep{burstGrbS5,cbcGrbS5}.
No evidence for a gravitational-wave signal was found in these searches.  
For GRB~051103 and GRB~070201, short-duration GRBs with position error boxes 
overlapping respectively the M81 galaxy at 3.6\,Mpc and the Andromeda galaxy 
(M31) at 770\,kpc, the non-detection of associated gravitational waves ruled 
out the progenitor object being a compact binary coalescence in M81 or M31 with high confidence 
\citep{grb070201_07,grb051103}.  

Although it is expected that most GRB progenitors will be at distances too 
large for the resulting gravitational-wave signals to be detectable by LIGO and Virgo  
\citep{berger05}, it is possible that a few GRBs could be located nearby.
For example, the smallest observed redshift to date of an optical GRB afterglow 
is $z=0.0085$ ($\simeq36$ Mpc) for GRB~980425 \citep{galama98,kulkarni98,iwamoto98}; 
this would be within the LIGO-Virgo detectable range for some progenitor models.  
Recent studies \citep{NatureSoderberg2006,2007MNRAS.382L..21C,Le07,2007ApJ...662.1111L,Virgili07} indicate the 
existence of a local population of under-luminous long GRBs with an observed 
rate density 
approximately $10^3$ times that of the high-luminosity population.
Also, observations suggest that short-duration GRBs tend to have smaller 
redshifts than long GRBs \citep{GuPi:05,fox05}, and this has led to fairly 
optimistic estimates \citep{CBCrate,Leonor09} for detecting associated 
gravitational-wave emission.
Approximately 90\% of the GRBs in our sample do not have measured redshifts,
so it is possible that one or more could be much closer than the typical $\sim$Gpc 
distance of GRBs.  

In this paper, we present the results of a search for gravitational waves 
associated with \nAnalyzedGRB\ GRBs that were detected by 
satellite-based gamma-ray experiments during the sixth LIGO science run 
and second and third Virgo science runs, which collectively 
spanned the period from 2009 July 7 to 2010 October 20. 
We search for coalescence signals associated with 26 short GRBs and unmodeled 
GW bursts associated with \nBurstGRB\ GRBs (both short and long). 
The search for unmodeled GW bursts targets signals with duration $\lesssim1$ s  
and frequencies in the most sensitive LIGO/Virgo band, approximately 
60 Hz $-$ 500 Hz. 
We find no evidence for a gravitational-wave candidate associated with any
of the GRBs in this sample, and statistical analyses of the GRB sample 
show no sign of a collective signature of weak gravitational waves. 
We place lower bounds on the distance to the progenitor for each GRB, 
and constrain the fraction of the observed GRB population at low redshifts. 

The paper is organized as follows.  Sec.~\ref{sec:signal} discusses 
the GW signal models that are used in these searches.  Sec.~\ref{sec:s6run} 
briefly describes the LIGO and Virgo gravitational-wave detectors.  
Sec.~\ref{sec:grbsample} describes the GRB sample during the
2009-2010 LIGO-Virgo science runs, 
and Sec.~\ref{sec:search} summarizes the analysis procedure for GW burst
signals and for coalescence signals.
The results are presented in Sec.~\ref{sec:results} and
discussed in Sec.~\ref{sec:discussion}.  We conclude
in Sec.~\ref{sec:conclusion} with some comments on the astrophysical
significance of these results and the prospects for GRB searches in
the era of advanced gravitational-wave detectors.

\section{GW signal models}
\label{sec:signal}

As noted above, the progenitors of long GRBs are 
extreme cases of stellar collapse, while the most plausible progenitors 
of the majority of short GRBs are mergers of a neutron star with either 
another neutron star or a black hole.  In this section we review the 
expected GW emission associated with each scenario, and the expected 
delay between the gamma-ray and GW signals.

\subsection{GWs from extreme stellar collapse}
\label{sec:signal:longGRB}

Stellar collapse is notoriously difficult to model. It necessitates
complex micro-physics and full three-dimensional simulations, which
take years to complete for a single initial state. Many simulations
that include some, but not all, physical aspects have been performed for
non-extreme cases of core-collapse supernovae, which identified
numerous potential GW burst
emission channels; see \cite{Ott09-1} for a review. These models predict emission of up
to $10^{-8}\,\mathrm{M_\odot c^2}$ through GWs. Given the sensitivity of
current GW detectors, such GW emission models are not detectable from
extra-galactic progenitors.

However, in the extreme stellar collapse conditions which are
necessary to power a GRB, more extreme GW emission channels can be
considered. Several semi-analytical scenarios have been proposed which
produce up to $10^{-2}\,\mathrm{M_\odot c^2}$ in GWs,
all of which correspond to some rotational instability developing in the
GRB central engine \citep{Davies02,Fryer02,Kobayashi03,Shibata03,Piro07,Corsi09,Romero10}. 
In each model the GWs are emitted by a quadrupolar mass
distribution rotating around the GRB jet axis.
Given the observation of a GRB, this axis is roughly pointing at the
observer, which yields circularly polarized GWs \citep{Kobayashi03-1}. 

For extreme stellar collapses, the arrival of $\gamma$-rays can be
significantly delayed with respect to the GW emission. Delays of up to
100\,s can be due to several phenomena: the delayed emission of the
relativistic jet \citep{MacFadyen01}; 
sub-luminal propagation of the jet to the surface of the star in the
collapsar model for long GRBs \citep[see for
example][]{AlMuIbMaMa:00,ZhWoMa:03,WaMe:07,LaMoBe:09}; and the duration, in
the observer's frame, of the
relativistic propagation of the jet before the onset of the prompt $\gamma$-ray
emission  \citep{Vedrenne:ch5}. For some GRBs, $\gamma$-ray precursors have
been observed up to
several hundred seconds before the main $\gamma$-ray emission
peak \citep{1995ApJ...452..145K,Burlon09,Burlon08,Lazzati05}; the precursor could mark the
initial event, with the main
emission following after a delay.

\subsection{GWs from a compact binary progenitor}
\label{sec:signal:shortGRB}

The coalescence of two compact objects is usually thought of as a three-step
process: an inspiral phase, where the orbit of the binary slowly
shrinks due to the emission of GWs; a merger phase, when the 
two objects plunge together; and a ringdown phase, during which the newly 
created and excited black hole settles into a stationary state~\citep{lrr-2011-6}. 
As the gravitational
waves emitted in the inspiral phase dominate the signal-to-noise-ratio 
in current detectors, we focus on that phase only.\footnote{
For high-mass systems the merger and ringdown phases can contribute significantly
to the signal-to-noise-ratio with current detectors. However, given the mass range used in GRB searches,
the merger and ringdown phases can be ignored.} 

We consider compact binaries consisting of two neutron stars (NS-NS) or a
neutron star with a black hole (NS-BH).  As the objects spiral
together, the neutron star(s) are expected to tidally disrupt shortly
before they coalesce, creating a massive torus.  The matter in the torus can
then produce highly relativistic jets, which are supposedly ejected along the axis of
total angular momentum.  While this picture is supported by recent 
numerical simulations \citep{Foucart11, Rezzolla11}, it has not yet been 
confirmed by complete simulations, and the influence of a tilted
BH spin is uncertain.

Contrary to the long GRB case, the onset of $\gamma$-ray emission is
delayed only up to a few seconds compared to the GW emission, as there
is no dense material retaining the jet and other delay effects are at
most as long as the GRB duration \citep{Vedrenne:ch5}.
Semi-analytical calculations of the
final stages of a NS--BH coalescence show that the majority of matter
plunges onto the BH within $\sim$\unit{1}{s} \citep{Davies:2004pu}.
Numerical simulations of the mass transfer suggest a timescale of
milliseconds or a few seconds at maximum
\citep{Faber:2006tx,Rosswog:2006ue,Etienne:2007jg,Shibata:2007zm}.
Therefore, an observer in the cone of the collimated outflow 
is expected to observe the gravitational-wave signal up to a few seconds 
before the electromagnetic signal from the prompt emission.

\section{LIGO Science Run 6 \& Virgo Science Runs 2-3}
\label{sec:s6run}

The LIGO and Virgo detectors are kilometer-scale, power-recycled 
Michelson interferometers with orthogonal Fabry-Perot arms 
\citep{abbottnim04,abbott-2007,1748-0221-7-03-P03012}.  They are designed to 
detect gravitational waves with frequencies ranging from 
$\sim40$\,Hz to several kHz, with maximum 
sensitivity near 150\,Hz.  There are two LIGO observatories: one 
located at Hanford, WA and the other at Livingston, LA.  The Hanford site
houses two interferometers: one with 4\,km arms (H1) and the other with 2 km
arms (H2). The Livingston observatory has one 4\,km interferometer (L1).
The two
observatories are separated by a distance of 3000\,km, corresponding to a 
travel time of 10\,ms for light or gravitational waves.
The Virgo detector (V1) is in Cascina near Pisa, Italy.  
The time-of-flight separation between the Virgo and Hanford observatories 
is 27\,ms, and between Virgo and Livingston is 26\,ms.

A gravitational wave is a spacetime metric perturbation that is manifested 
as a time-varying quadrupolar strain, with two polarization components.
Data from each interferometer records the length difference of the 
arms and, when calibrated, 
measures the strain induced by a gravitational 
wave.  These data are in the form of a time series, digitized at a 
sample rate of 16384\,Hz (LIGO) or 20000\,Hz (Virgo).  

\begin{figure}
\includegraphics[width=0.45\textwidth]{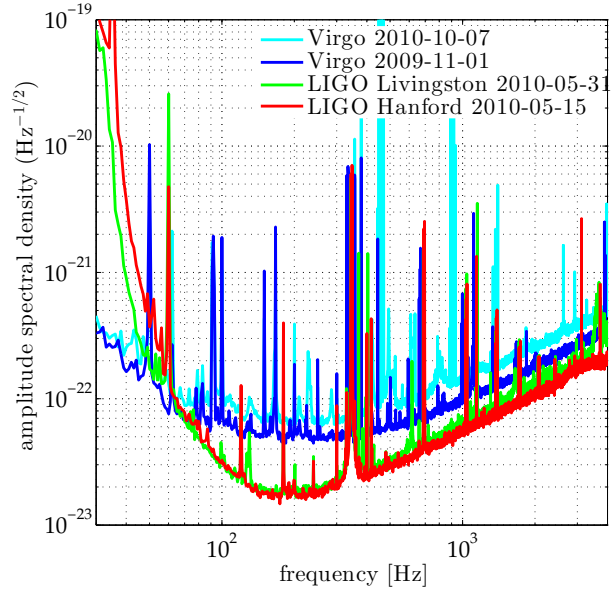}
\caption{\label{fig:spectra} Best strain noise spectra from the LIGO 
and Virgo detectors during the 2009-2010 science runs.}
\end{figure}

The sixth LIGO science run was held from 2009 July 07 to 2010 October 20.  
During this run, the 4 km H1 and L1 detectors were operated at sensitivities 
that surpassed that of the previous 2005-2007 run, with duty factors of 52\% and 47\%.  
The 2 km H2 detector was not operated in 2009-2010. 
The second Virgo science run was held from 2009 July 07
to 2010 Jan 08 with an improvement in sensitivity roughly a factor
of 2 over Virgo's first science run. 
The third Virgo science run was held from 2010 Aug 11
to 2010 Oct 20.
The overall Virgo duty cycle over these two science runs was 78\%. 
Fig.~\ref{fig:spectra} shows the best sensitivities, in terms of 
noise spectral density, of the LIGO and Virgo interferometers during these runs.
The distance at which the LIGO instruments would observe an optimally oriented,
optimally located coalescing neutron-star binary system with a signal-to-noise-ratio of 8
reached about 40\,Mpc; for Virgo the same figure of merit reached about 20\,Mpc.

The GEO~600 detector \citep{geo08}, located near Hannover, Germany, was also 
operational in 2009-2010, though with a lower sensitivity than LIGO 
and Virgo. We do not use the GEO data in this search as the 
modest gains in the sensitivity to gravitational-wave signals would not have 
offset the increased complexity of the analysis.  
However, GEO data is used in searches for gravitational 
waves coincident with GRBs occurring during periods when only one of
the LIGO or Virgo detectors is operational, such as the period between
the fifth and sixth LIGO science runs and during summer 2011. The result of
those searches will be reported in a future publication.

\section{GRB sample}
\label{sec:grbsample}

We obtained our sample of GRB triggers from the Gamma-ray
burst Coordinates
Network\footnote{\href{http://gcn.gsfc.nasa.gov/}{http://gcn.gsfc.nasa.gov/}}
(GCN) \citep{Barthelmy08}, supplemented by the
\emph{Swift}\footnote{\href{http://gcn.gsfc.nasa.gov/swift_gnd_ana.html}{http://gcn.gsfc.nasa.gov/swift\_gnd\_ana.html}} and
\emph{Fermi}\footnote{\href{http://heasarc.gsfc.nasa.gov/W3Browse/fermi/fermigtrig.html}{http://heasarc.gsfc.nasa.gov/W3Browse/fermi/fermigtrig.html}}
trigger pages.  This sample of GRB triggers 
came mostly from the {\em Swift} satellite \citep{swift04} and the
\emph{Fermi} satellite \citep{GBM09}, but several triggers also came from
other spaceborne experiments, such as MAXI \citep{MAXI09},
SuperAGILE \citep{superAGILE07} and INTEGRAL
\citep{integral03}, as well as from time-of-flight triangulation using
satellites in the third InterPlanetary network (IPN) \citep{ipn2009}.

In total there are \nTotGRB\ GRBs in our GRB sample
during the 2009-2010 LIGO-Virgo science runs.
About 10\% of the GRBs have associated redshift measurements, 
all of them evidently beyond the reach of current GW detectors.  Nevertheless, 
times around these GRBs have been analyzed in case of, for example, a 
chance association with an incorrect host galaxy.

GRBs that occurred when two or more of the LIGO and Virgo detectors 
were operating in a resonant and stable configuration are analyzed.  
Data segments which are flagged as being of poor quality are 
excluded from the analysis. In total, \nAnalyzedGRB\ GRBs were
analyzed, out of which \nBurstGRB\ GRBs were
analyzed by the GW burst search, and 26 short GRBs were analyzed by
the coalescence search. (As the GW data quality requirements are
somewhat different for the unmodeled burst and coalescence searches, 4
short GRBs analyzed by the coalescence search could not be analyzed by
the GW burst search.)

The classification of GRBs into short and long is somewhat ambiguous 
\citep{Gehrels:2006tk,Bloom:2008cn, Zhang:2009uf, Horvath:2010um}.
Since binary mergers are particularly strong sources
of gravitational radiation, we make use of a more lenient classification
to identify GRBs which may originate from a binary merger  
\citep{Zhang:2006mb,Zhang:2009uf}.  Our selection is based on the 
$T_{90}$ duration (the time interval over which 90\% of
the total background-subtracted photon counts are observed), and on 
visual inspection of all available lightcurves. 
Specifically, we treat as ``short'' all GRBs with $T_{90} < 4$\,s; this choice, rather 
than the standard 2\,s cutoff for short GRBs, is to ensure we include 
those short GRBs in the tail of the duration distribution.
In addition, some of the
longer-duration GRBs exhibit a prominent short spike at the beginning of the
lightcurve and an extended longer emission \citep{Norris06}, suggesting that
those GRBs might be created by the merger of two compact objects.
Those GRBs were also treated as short GRBs and, where necessary, the 
trigger time used for the coalescence search was shifted by up to a few seconds 
to match the rising edge of the spike (which should correspond to the 
binary coalescence time).  This lenient classification ensures a relatively 
complete sample, at the price of sample purity -- some of the GRBs we analyze
as ``short'' may not have a compact binary progenitor. This impurity is acceptable
for the purpose of GW detection where we do not want to miss a potentially
observable GW counterpart.
The final set of 26 short GRBs is given in Tab.~\ref{tab:shortGRB}.

A large number of GRBs detected by the IPN are not reported by the
Gamma-ray burst Coordinates Network; the result of a search for GWs associated with those GRBs will be
reported in a future publication.

\section{Searches for GWs associated with GRBs}
\label{sec:search}

We perform searches for both unmodeled bursts and coalescence signals. 
We begin this section by describing the basic methodology and features 
common to both searches, then briefly present the details of the two 
analysis methods.

\subsection{Search Methodology}
\label{sec:method}

Both search pipelines identify an ``on-source''  time in which to
search for an associated GW event. This time selection is expected to
improve by a factor $\sim$1.5 the sensitivity of the search compared to an
all-sky / all-time search \citep{kochanek93}.  
For the GW burst search, we use the 
interval from 600\,s before each GRB trigger to either 60 s or the 
$T_{90}$ time (whichever is larger) after the trigger as the window
in which to search for a GW signal.  This conservative window is large
enough to take into account most plausible time delays between a
GW signal from a progenitor and the onset of the gamma-ray signal, 
as discussed in Sec.~\ref{sec:signal:longGRB}.  This window is
also safely larger than any uncertainty in the definition of the
measured GRB trigger time.  For cases when less early GW data are available,
a shorter window starting 120\,s before the GRB trigger time is used. This
still covers most time-delay scenarios. 
For the binary coalescence search, it is believed that the delay between 
the merger and the emission of $\gamma$-rays will be small, as
discussed in Sec.~\ref{sec:signal:shortGRB}. We therefore 
use an interval of 5 s prior to the GRB to 1 s following as the on-source
window, which is wide enough to allow for uncertainties in the emission 
model and in the arrival time of the electromagnetic signal \citep{cbcGrbS5}. 

The on-source data are scanned by the search algorithms to detect
possible GW transients (either coalescence or burst), referred to as ``events''. 
For both searches the analysis depends on the sky position of the GRB.  
GRBs reported by the \emph{Swift} satellite have very small position uncertainty 
\cite[$\ll 1^\circ$; see][]{Barthelmy:2005hs}, and 
the GW searches need only be performed at the reported sky location. 
For GRBs detected by the Gamma-ray Burst Monitor (GBM) on the \emph{Fermi}
satellite \citep{GBM09}, however, the sky localization region can be 
large ($\gg 1^\circ$), and detection efficiency would be lost
if the GW searches only used a single sky location. To resolve this 
problem, searches for poorly localized GRBs are done over a grid of sky
positions, covering the sky localization region \citep{thesisWas,Was12}. 
We assume a systematic 68\% coverage error circle for the Fermi/GBM sky
localizations with a radius of $3.2^\circ$ with 70\% probability and a
radius of $9.5^\circ$ with 30\% probability \citep{GBMsysErr_GCN},
which is added in quadrature to the reported statistical error.

Each pipeline orders events found in the on-source time according to
a ranking statistic. To reduce the effect of non-stationary
background noise, candidate events are subjected to checks
that ``veto'' events overlapping in time with known instrumental or
environmental disturbances \citep{VSR1-4DetChar}. The surviving
event with the highest ranking statistic is taken to be the best
candidate for a gravitational-wave signal for that GRB; it is referred
to as the {\it loudest event} \citep{Brady:2004gt,Biswas:2007ni}.  To
estimate the significance of the loudest event,
the pipelines also analyze coincident data from a period
surrounding the on-source data, where we do not expect a signal.
The proximity of this {\it off-source} data to the
on-source data makes it likely that the estimated background will
properly reflect the noise properties in the on-source segment.  The
off-source data are processed identically to the on-source data; in
particular, the same data-quality cuts and consistency tests are applied, 
and the same sky positions relative to the GW detector network are used. 
If necessary, to increase the background distribution statistics,
multiple time shifts are applied to the data streams from different
detector sites, and the off-source data re-analyzed for each time shift.  

To determine if a GW is present in the on-source data, the loudest
on-source event is compared to the distribution of loudest off-source 
events.  A p-value is defined as the probability 
of obtaining such an event or louder in the onsource, given the background 
distribution, under the null hypothesis.  The triggers with the 
smallest p-values in the searches are subjected to additional followup 
studies to determine if the events can be associated 
with some non-GW noise artifact, for example due
to an environmental disturbance.

Regardless of whether a statistically significant signal is present, we 
also set a 90\% confidence level lower limit on the distance to the GRB 
progenitor for various signal models. 
This is done by adding simulated GW signals to the data and repeating the
analyses. These signals, which are drawn from astrophysically motivated
distributions described in the following sections, are used to calculate 
the maximum distance for which there is a 90\% or greater chance that such
a signal model, if present in the on-source region, would have
produced an event with larger ranking statistic than the largest value
actually measured.

\subsection{Search for GW bursts}
\label{sec:burst}

The search procedure for GW bursts follows that used in the 2005-2007 GRB search
\citep{burstGrbS5}.  All GRBs are treated identically, without regard
to redshift (if known), fluence or classification.  The on-source data
are scanned by the \xp~algorithm \citep{Sutton:2009gi,Was12}, which is 
designed to detect short GW bursts, $\lesssim1$\,s, in the $60-500$\,Hz 
frequency range.
\xp~combines data from arbitrary sets of detectors, taking into
account the antenna response and noise level of each detector to
improve the search sensitivity.  Time-frequency maps of the combined 
data streams are scanned for clusters of pixels with energy significantly 
higher than that expected from background noise.  The resulting 
candidate GW events are characterized by a ranking statistic based on energy.
We also apply consistency tests based on the signal correlations
measured between the detectors, assuming a circularly polarized GW, 
to reduce the number of background events. (The circular polarization 
assumption is motivated by the fact that the GRB system rotation axis 
should be pointing roughly at the observer, as discussed in 
Sec.~\ref{sec:signal:longGRB}.) 
The stringency of these tests is tuned by comparing their effect on 
background events and simulated signal events. The
background samples are constructed using the $\pm1.5$~hours of data
around the GRB trigger, excluding the on-source
time. Approximately 800 time shifts of these off-source data
are used to obtain a large sample of background events.

To obtain signal samples, simulated signals are added to the on-source
data.
The models of GW emission by extreme stellar collapse described in
Sec.~\ref{sec:signal:longGRB} do not predict the exact shape
of the emitted GW signal. As an ad-hoc model, we use the GW emission by
a rigidly rotating quadrupolar mass moment with a Gaussian time
evolution of its magnitude. For such a source with a rotation axis
inclined by an angle $\iota$ with respect to the observer the received
GW signal is a sine-Gaussian
\begin{multline}\label{eq:ellSG}
  \begin{bmatrix}
    h_+(t) \\ h_\times(t) 
  \end{bmatrix} =
  \frac{1}{r}\sqrt{\frac{G}{c^3}\frac{E_\text{GW}}{f_0 Q} \frac{5}{4
      \pi^{3/2}}} \times \\
  \begin{bmatrix}
    (1+ \cos^2 \iota) \cos(2 \pi f_0 t) \\
    2\cos \iota \sin(2 \pi f_0 t)
  \end{bmatrix}\exp\left[ - \frac{(2 \pi f_0 t)^2}{2Q^2}\right],
\end{multline}
where the signal frequency $f_0$ is equal to twice the rotation
frequency, $t$ is the time relative to the signal peak time, 
$Q$ characterizes the number of cycles for which the
quadrupolar mass moment is large, $E_\text{GW}$ is the total
radiated energy, and $r$ is the distance to the source. 
We consider two sets\footnote{\xp~also uses
sine-Gaussian signals with $f_0=100$\,Hz and non-spinning coalescence
signals, as discussed in Sec.~\ref{sec:cbc}, to tune the
pipeline.} of
such signals with signal frequencies $f_0$ of 150\,Hz and 300\,Hz,
which covers the sensitive frequency band of this GW burst search. The
inclination angle is distributed uniformly in $\cos \iota$, with $\iota$
between $0^\circ$ and $5^\circ$, which corresponds to the typical jet
opening angle of $\sim 5^\circ$ observed for long GRBs
\citep{Gal-Yam06,Racusin09}.

Systematic errors are marginalized over in the sensitivity
estimation by ``jittering'' the simulated signals before adding them to
the detector noise. This includes distributing injections across the
sky according to the gamma-ray satellites' sky location error box, and
jittering the signal amplitude, phase, and timing in each detector
according to the given detector calibration errors 
\citep{VSR2calibration,s6calibration}. This procedure also ensures that 
the consistency tests used in the analysis are loose enough to allow
for such errors.

\subsection{Search for GWs from a compact binary progenitor}
\label{sec:cbc}

The core of the coalescence search involves correlating the measured data
against theoretically predicted waveforms using matched filtering 
\citep{helmstrom-1968}.
GWs from the inspiral phase of a coalescence are modeled by 
post-Newtonian approximants in the band of the detector's sensitivity 
for a wide range of binary masses \citep{Blanchet:2006av}. 
The expected GW signal depends on the masses and spins of the NS and 
its companion (either NS or BH), as well as the distance to the source, 
its sky position, its inclination angle, and the polarization angle 
of the orbital axis. 
Matched filtering is most sensitive to the phase evolution of the 
signal, which depends on the binary masses and spins, the time of 
merger, and a fiducial phase.  The time and phase can be determined 
analytically.  Ignoring spin, we can therefore perform 
matched filtering over a discrete two-dimensional bank of templates 
which span the space of component masses.  This bank is constructed 
such that the maximum loss in signal-to-noise ratio for a binary with
negligible spins is $3\%$ \citep{hexabank,Harry:2010fr}.
For this search, as in the 2005-2007 GRB search \citep{cbcGrbS5}, we used ``TaylorF2''
frequency domain templates, generated at 3.5 post-Newtonian order
\citep{Blanchet:1995ez,Blanchet:2004ek}.
While the spin of the components is ignored in the template waveforms, 
we evaluate the efficiency of the search using 
simulated signals including spin, as described below.

For each short GRB, the detector data streams are combined coherently and 
searched using the methods described in detail in \citet{Harry:2010fr}.
Various signal consistency tests are then applied
to reject non-stationary noise artefacts.  These include
$\chi^2$ tests \citep{Allen:2004gu, Hanna:2008}, a null stream
consistency test, and a re-weighting of the signal-to-noise-ratio to take into account the
values recorded by these tests. 
This is the first coherent search for coalescence signals; it has been found 
to be more sensitive to GW signals than the coincidence technique used 
in previous triggered coalescence searches of LIGO and Virgo data \citep{Harry:2010fr}.  
Tests using the simulations described below have also shown that 
this focused coalescence search is a factor of $\sim$2 more sensitive to coalescence signals than the 
unmodeled search described in the previous section, justifying the use 
of a specialized search for this signal type.

To estimate the efficiency of the search and calculate exclusion 
distances for short GRBs, we draw
simulations from two sets of astrophysically motivated compact binary
systems: two neutron stars (NS--NS); and a 
neutron star with a black hole (NS--BH).  
The NS masses are chosen from a Gaussian distribution
centered at 1.4\,$\mathrm{M_\odot}$ \citep{Kiziltan:2010ct,Ozel:2012ax}
 with a width of
0.2\,$\mathrm{M_\odot}$ for the NS--NS case, and a broader spread of
0.4\,$\mathrm{M_\odot}$ for the NS--BH systems, to account for larger
uncertainties given the lack of observations for such systems.  
The BH masses are Gaussian distributed with a mean of
10\,$\mathrm{M_\odot}$ and a width of~6\,$\mathrm{M_\odot}$. The BH mass is
restricted such that the total mass of the system is less than $25\,M_{\odot}$.
For masses greater than this, the NS would be `swallowed whole' 
by the BH, no massive torus would form, and no GRB would be produced
\citep{0264-9381-27-11-114002,Ferrari:2009bw,lrr-2011-6}.

Observed pulsar spin periods and assumptions about the spindown rates
of neutron stars 
place the NS spin periods at birth in the range of 10 to 140\,ms,
corresponding to an upper limit on $S / m^2$ of
$\leq 0.04$ \citep{Mandel:2009nx}, where $S$ denotes the spin
of the neutron star and $m$ its mass.  However, neutron stars can be spun up
to much higher spins (e.g.\ to 716\,Hz \citep{Hessels:2006ze}), hence 
we conservatively assume a maximum spin of $S / m^2<0.4$ corresponding
to a $\sim$1 ms pulsar.  Therefore, the spin magnitudes are drawn
uniformly from the range $[0,\,0.4]$.  For BHs
the magnitudes are chosen uniformly in the $[0,\,0.98)$ range
\citep{Mandel:2009nx}.
The spins are oriented randomly, with a constraint on the
tilt angle (the angle between the spin direction of the BH and
the orbital angular momentum).  Since the merger needs to power a GRB, a
sufficiently massive accretion disk around the BH is required.
Population synthesis studies indicate that the tilt angle is
predominantly below 45$^\circ$ \citep{Belczynski:2007xg}; numerical simulations show
that for tilt angles larger than 40$^\circ$ the mass of the disk will
drop rapidly \citep{Foucart11} and BHs with tilt angle $>60^\circ$
will `swallow' the NS completely, leaving no accretion disk to power a
GRB \citep{Rantsiou:2007ct}. In our simulations we use the weakest of
these three constraints and set the tilt angle 
to be $<60^\circ$.

The outflow from a GRB is most likely to be along the direction of the 
total angular momentum $J$ of the system as discussed in
Sec.~\ref{sec:signal:shortGRB}. 
Observations suggest that this outflow is confined within a cone, whose half-opening angle is estimated to range between several degrees to over 60$^\circ$ for short GRBs
\citep[see e.g.][]{Burrows:2006ar,Grupe:2006uc,Dietz:2010eh}. 
Under the assumption that this cone is centered along the total angular momentum $J$ of the system, we chose the inclination angle between $J$
and the line-of-sight to the observer to be distributed within cones
of half-opening angles 10$^\circ$, 30$^\circ$, 45$^\circ$ and 90$^\circ$. 
The majority of the results quoted in this work assume a 30$^\circ$ angle.

The
coalescence time is uniform over the on-source region, and the sky 
position of the GRB is jittered according to the
reported uncertainty of the location.

The quoted exclusion distances are marginalized over systematic errors 
that are inherent in this analysis. First, there is some uncertainty
in how well our PN templates will match real GW signals; we expect a 
loss in signal-to-noise-ratio of up to 10\% because of this mis-match \citep{Abbott:2009tt}.
Second, there is uncertainty in the amplitude calibration of the detectors
\citep{s6calibration,VSR2calibration}; phase and timing calibration uncertainties are also present, but are negligible compared to other sources of errors.

An opportunistic search for coalescence signals has also been performed on the long GRBs.
This search is done to conservatively account for uncertainties in the details of
the short/long GRB classification, and for uncertainties in the progenitor model
of long GRBs for which an associated SN signature was excluded 
\citep{Gehrels:2006tk,Watson:2007nk,Gao:2010yh}. We use the same
analysis to check for a coalescence signal associated with long GRBs, but do 
not estimate exclusion distances as the compact binary coalescence progenitor model is unlikely for long GRBs.

\subsection{Significance of p-value distribution}
\label{sec:fappop}

In addition to evaluating individual p-values,
we use a weighted binomial test to assess whether the obtained
set of p-values is compatible with the uniform distribution expected 
from noise only, for both the GW burst and coalescence searches. This test 
looks for deviations from the null hypothesis in the 5\% tail of 
lowest p-values weighted by the prior probability of detection 
(estimated from the GW search sensitivity).  
The weighted binomial test is an extension of the 
binomial test that has been used in previous searches for GW bursts 
associated with GRBs \citep{burstGrbS234,burstGrbS5}. 
The combination of p-values with prior detection probabilities gives 
more weight to GRBs for which the GW detectors had better sensitivity 
and therefore the detection of a GW signal is more likely.
The details of this test 
are given in Appendix~\ref{sec:weightedBinomial}.

The result of the weighted binomial test is a single ranking statistic 
$S_\text{weighted}$.  The statistical significance of the measured 
$S_\text{weighted}$ is assessed by comparing to the background 
distribution of this statistic from Monte Carlo simulations with p-values 
uniformly distributed in $[0,1]$. This yields the overall background 
probability of the measured set of p-values.

\section{Results}
\label{sec:results}

The coalescence analysis has been applied to search for signals in coincidence
with 26 short GRBs; the GW burst analysis has been applied to \nBurstGRB\
GRBs, which include 22 of the 26 short GRBs analyzed by the coalescence
search.  (As mentioned in Sec.~\ref{sec:grbsample}, 4 of the short
GRBs analyzed by the coalescence search could not be analyzed by the
GW burst
search.) The lists of analyzed GRBs classified as short and long are
given in Tab.~\ref{tab:shortGRB} and Tab.~\ref{tab:longGRB}.

\subsection{GW burst search results}

\begin{figure}
  \centering
  \includegraphics[width=0.45\textwidth]{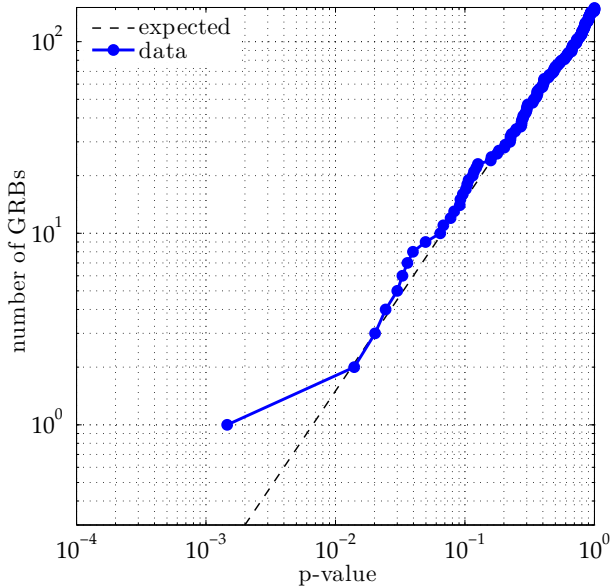}
  \caption{Cumulative p-value distribution from
    the analysis of \nBurstGRB\ GRBs with the GW burst search. The expected
    distribution under the null hypothesis is indicated by the
    dashed line. 
    }
  \label{fig:binomialTestBurst}
\end{figure}

The distribution of p-values for each of the  \nBurstGRB\ GRBs
analyzed by the GW burst search 
is shown in
Fig.~\ref{fig:binomialTestBurst}. 
The weighted binomial test yields a background probability of
25\%.  Therefore, the distribution is consistent with no 
GW events being present.

The smallest p-value, 0.15\%, has been obtained for GRB~100917A.
This GRB was localized on the sky by \emph{Swift},
however no redshift measurement is available to date. The
corresponding GW event was obtained by combining data from the
H1, L1, and V1 detectors. A study of the environmental and instrumental
channels at that time yields potential instrumental causes
for this event, but is not conclusive.
Regardless, the measured p-value is not significant as determined by the 
weighted binomial test, so this event is not a candidate for a
gravitational-wave detection.

\begin{figure}
  \centering
  \includegraphics[width=0.45\textwidth]{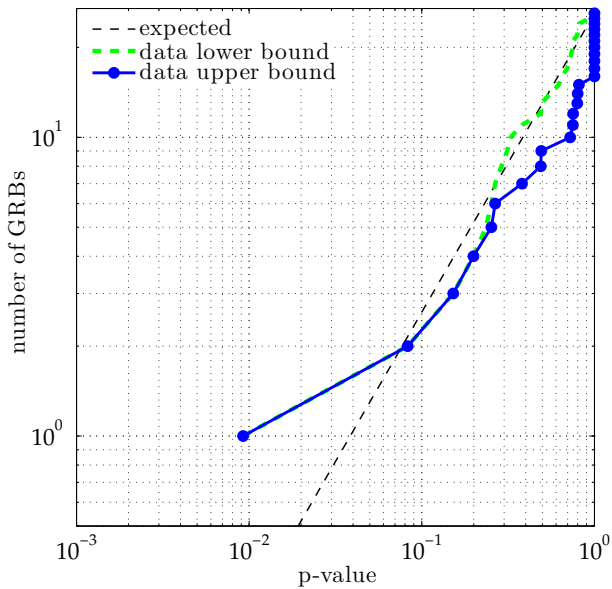}
  \caption{Cumulative p-value distribution from
    the analysis of 26 short GRBs with the coalescence search. For GRBs where no event
    is observed in the on-source region, we can only place a lower bound on the p-value,
    thus we show two distributions where 
    the upper (blue solid line) and lower (green dashed line) bound 
    respectively was taken for every GRB. The expected
    distribution under the null hypothesis is indicated by the
    dashed line.}
  \label{fig:binomialTestCBC}
\end{figure}

\subsection{Coalescence search results}

The distribution of p-values for each of the 26 short GRBs analyzed by the coalescence search is shown in 
Fig.~\ref{fig:binomialTestCBC}. 
The result of the weighted binomial  test yields a background probability of 8\%, corresponding to
a 1.8-sigma deviation from the null hypothesis. However, as we mentioned in section \ref{sec:grbsample},
we use a lenient classification when deciding if GRBs are treated as short or long for the purposes of
our analyses. If restrict our short GRB sample to the more commonly used criterion, 
$T_{90} < 2$\,s, then we find a background probability of 3\%, corresponding to a 2.2-sigma deviation.

This deviation was due to an event found in coincidence with
GRB~100328A, which produced the smallest p-value of 1\%, and was the GRB
to which the search had the second best sensitivity. A followup
investigation of this candidate determined that it was due to a noise artifact
in the Hanford instrument, which was one of a class of glitches caused
by a bad power supply which contaminated the length and angular
control servos. No other noteworthy events were found by this search
and thus there are no potential gravitational-wave candidates.  The
opportunistic search for coalescence signals associated with long GRBs did not
yield any candidate that was inconsistent with background noise.

\section{Astrophysical interpretation}
\label{sec:discussion}

Given that no significant event was found in our analyses, we place
limits on GW emission based on the signal models discussed in
Sec.~\ref{sec:signal}, and assess the potential of a similar
search with second-generation gravitational-wave detectors.

\subsection{Distance exclusion}

For each GRB we derive a 90\% confidence lower limit on the GRB
progenitor distance for various emission models 
using the methodology described in Sec.~\ref{sec:method}.

\begin{figure}
  \centering
  \includegraphics[width=0.45\textwidth]{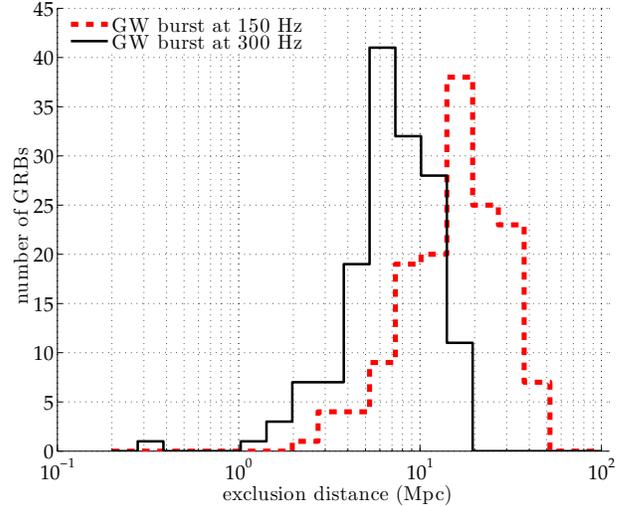}
  \caption{Histograms across the sample of GRBs of the distance 
    exclusions at the 90\% confidence level for circularly polarized 
    sine-Gaussian GW burst models at 150\,Hz and 300\,Hz. We assume
    an optimistic standard siren GW emission of 
    $E_\text{GW} = 10^{-2}\,\mathrm{M_\odot c^2}$. 
    See Tab.~\ref{tab:shortGRB} and Tab.~\ref{tab:longGRB} for the 
    exclusion values for each GRB.}
  \label{fig:dist_CSG}
\end{figure}

The GW burst search provides lower limits on the generic GW burst signal emitted 
by a rotator described in Sec.~\ref{sec:burst} for each GRB.
We assume that the source emitted 
$E_\text{GW} = 10^{-2}\,\mathrm{M_\odot c^2}$ 
of energy in gravitational waves
\footnote{We assume here an astrophysical model of a rotator which
  emits GWs mainly along the rotation axis. In previous searches
  \citep{burstGrbS5,grb070201_07} an unphysical isotropic GW emission
  of circularly polarized GWs was used. This change in model
  increases the distance exclusions presented here by a factor
  $\sqrt{5/2}$ relative to previous searches.},
that the jet opening angle is $5^\circ$,
and consider emission frequencies of 150\,Hz and 300\,Hz. The distance 
limits are given in Tab.~\ref{tab:shortGRB} and Tab.~\ref{tab:longGRB}, 
and their histogram is shown in Fig.~\ref{fig:dist_CSG}. The median 
exclusion distance is 
$D \sim 17\,\mathrm{Mpc} \, (E_\text{GW}/10^{-2}\,\mathrm{M_\odot c^2})^{1/2}$ 
for emission at frequencies around 150\,Hz, where the
LIGO-Virgo detector network is most sensitive.

\begin{figure}
  \centering
  \includegraphics[width=0.45\textwidth]{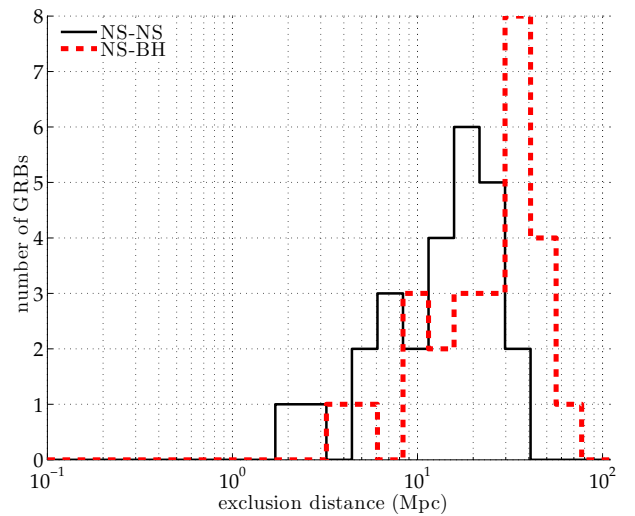}
  \caption{Histograms across the sample of short GRBs of the
    distance exclusions at the 90\% confidence level for
    NS--NS and NS--BH systems. See Tab.~\ref{tab:shortGRB} 
    for the exclusion values for each short GRB.}
  \label{fig:dist_insp}
\end{figure}

\begin{figure}
  \centering
  \includegraphics[width=0.45\textwidth]{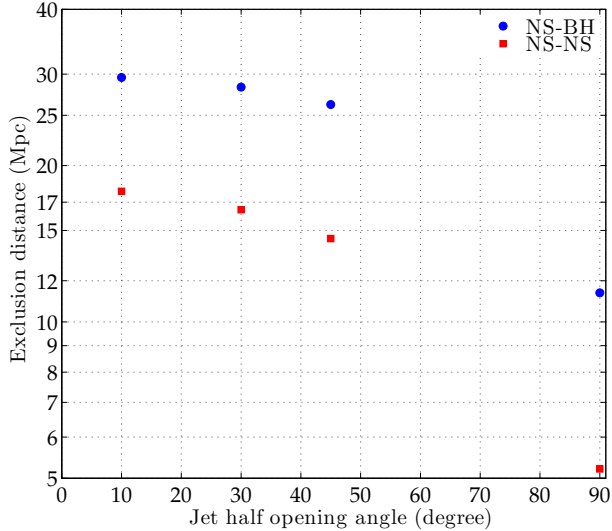}
  \caption{Median exclusion distances of compact binary coalescence sources as a function of
    half-opening angle, sampled at 10$^\circ$, 30$^\circ$, 45$^\circ$, 
    and 90$^\circ$.  The medians are computed over the set of 26 short GRBs, 
    for both NS-NS and NS-BH, at 90\% confidence level.}
  \label{fig:dist_excl_opening}
\end{figure}

The coalescence search sets lower limits on both the NS-NS and NS-BH models 
described in Sec.~\ref{sec:cbc} for each short GRB, assuming a jet 
half-opening angle of 30$^\circ$. The distance limits are given in 
Tab.~\ref{tab:shortGRB} and a
histogram of their values is shown in Fig.~\ref{fig:dist_insp}. The
median exclusion distance for NS--NS (NS--BH) coalescences is
\distNSNS\,Mpc (\distNSBH\,Mpc) for the 30$^\circ$ cone.  
We note that these exclusion distances are affected by the choice of 
signal parameter priors in Sec.~\ref{sec:cbc}; for example, 
Fig.~\ref{fig:dist_excl_opening} shows the median exclusion distances 
for half-opening angles of 10$^\circ$, 30$^\circ$, 45$^\circ$, and 90$^\circ$.  
Since the 
amplitude of a GW signal is stronger for a face-on binary, the
exclusion distance improves for smaller half-opening angles.  With no
restriction on the opening angle, the 90\% exclusion distance decreases
significantly, as there are orientations which will give very little
observable GW signal in the detector network.   

The GW burst and compact binary coalescence exclusion distances may be compared to those  
from all-sky searches, which look for GWs without requiring 
association with a GRB or other external trigger.  
Figure 7 of \citet{2012arXiv1202.2788T} presents 50\%-confidence 
exclusion energies for the all-sky GW burst search on this same data 
set for an assumed source distance of 10\,kpc, with a best limit of 
approximately $E_\text{GW} = 2 \times 10^{-8}\,\mathrm{M_\odot c^2}$ at 150 Hz.  
Rescaling to our nominal value of $10^{-2}\,\mathrm{M_\odot c^2}$ 
gives an exclusion distance of $\sim$7 Mpc.  
\citet{thesisWas} performs a more rigorous comparison that accounts 
for the fraction of events that do not produce GRB triggers due to 
the $\gamma$-ray beaming.  This indicates that for emission opening 
angles in the $5 - 30^\circ$ range, the GRB triggered search should 
detect a similar number of GW events coming from GRB progenitors as 
that detected by the all-sky search -- between 0.1 and 6 times the 
number detected by the all-sky search.  Furthermore, 
most of the GW events found by the GRB triggered search will be new 
detections not found by the all-sky search, illustrating the value 
of GRB satellites for gravitational-wave detection.

The NS--NS / NS--BH models used for compact binary coalescence exclusions stand 
on much firmer theoretical ground than the model used for GW burst exclusions.
The amplitude of a coalescence signal is well known and depends on the masses and spins
of the compact objects whereas the GW burst energy emitted
during a GRB is largely unknown and could be orders of magnitude
smaller than the chosen nominal value of $E_\text{GW} =
10^{-2}\,\mathrm{M_\odot c^2}$. In the pessimistic scenario that GRB
progenitors have a comparable GW emission to core-collapse supernova, 
the emitted energy could be as low as $E_\text{GW} \sim
10^{-8}\,\mathrm{M_\odot c^2}$. Such a signal would only be
observable with current gravitational wave detectors from a galactic source.

\subsection{Population exclusion}

As well as a per-GRB distance exclusion, we set an exclusion on GRB
population parameters by combining results from the set of analyzed
GRBs.  To do this, we use a simple population model, where all GRB
progenitors have the same GW emission (standard sirens), and perform
exclusion on cumulative distance distributions. We parametrize the
distance distribution with two components: a fraction $F$ of GRBs
distributed with a constant co-moving density rate\footnote{
While the distribution of the electromagnetically {\it observed} GRBs 
which serve as our triggers needs not 
be uniform in volume, this is a reasonable approximation at the 
distances to which LIGO-Virgo are sensitive. 
} 
up to a luminosity distance $R$, and a fraction
$1-F$ at effectively infinite distance. This simple model yields a
parameterization of astrophysical GRB distance distribution models that
predict a uniform local rate density and a more complex dependence at
redshift $>0.1$, as the large redshift part of the distribution is well
beyond the sensitivity of current GW detectors.  The exclusion is then
performed in the $(F,R)$ plane. Full details of the exclusion method are
given in Appendix~\ref{sec:appendix:popExclMethod}.

\begin{figure}
  \includegraphics[width=0.45\textwidth]{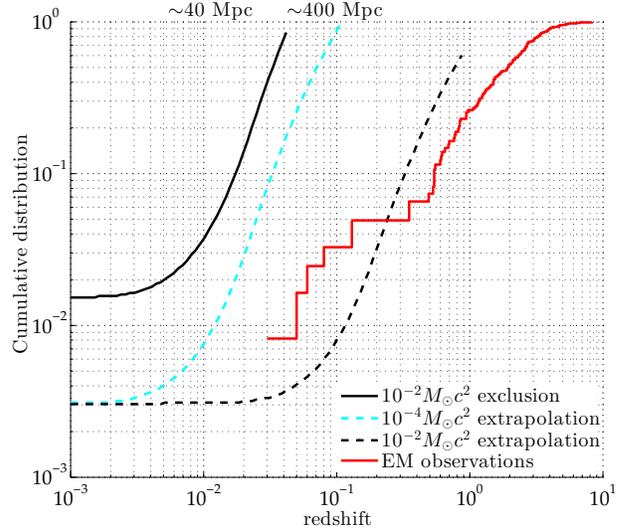}
  \caption{
    Cumulative redshift distribution $F(R)$ exclusion from the analysis of 
    \nBurstGRB\ GRBs with the GW burst search. We exclude at 90\% confidence level 
    cumulative distance distributions which pass through the region 
    above the black solid curve. We assume a standard siren 
    sine-Gaussian GW burst at $150\,\mathrm{Hz}$ with an
    energy of $E_\text{GW}=10^{-2}\,\mathrm{M_\odot c^2}$.  
    We extrapolate this exclusion to Advanced LIGO/Virgo assuming a
    factor 10 improvement in sensitivity and a factor 5 increase in
    number of GRB triggers analyzed. The black dashed curve is the
    extrapolation assuming the same standard siren energy of
    $E_\text{GW}=10^{-2}\,\mathrm{M_\odot c^2}$ and the cyan (gray)
    dashed curve assuming a less optimistic standard siren energy of
    $E_\text{GW}=10^{-4}\,\mathrm{M_\odot c^2}$
    \citep{Ott06,Romero10}. For reference, the red staircase curve
    shows the cumulative distribution of measured redshifts for 
    \emph{Swift} GRBs~\citep{Jakobsson06,Jakobsson12}.
  }
  \label{fig:zExcl_csg150}
\end{figure}

\begin{figure}
  \centering
  \includegraphics[width=0.45\textwidth]{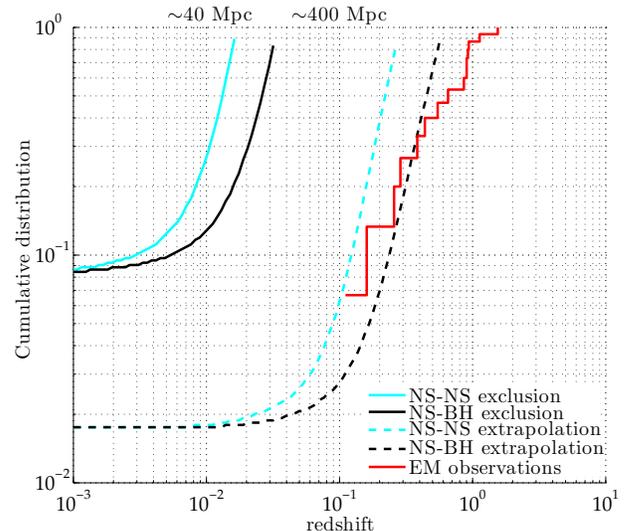}
  \caption{
    Cumulative redshift distribution $F(R)$ exclusion from the analysis of 
    26 short GRBs with the coalescence search.  Assuming that all the analyzed 
    short GRBs are NS--BH mergers (NS--NS mergers), we exclude at 90\% 
    confidence level cumulative distance distributions which pass through 
    the region above the black solid curve (cyan solid curve). 
    The dashed curves are the
    extrapolation of the solid curves to Advanced LIGO/Virgo, assuming
    a factor 10 improvement in sensitivity and a factor 5 increase in
    number of GRB triggers analyzed.
    For reference, the red staircase curve shows the cumulative
    distribution of measured redshifts for short GRBs~\citep{Dietz:2010eh}.
  }
  \label{fig:zExcl_cbc}
\end{figure}

The exclusion for GW bursts at 150\,Hz with
$E_\text{GW}=10^{-2}\,\mathrm{M_\odot c^2}$ is shown in
Fig.~\ref{fig:zExcl_csg150}, whereas the exclusion for the coalescence model
for short GRBs is shown in Fig.~\ref{fig:zExcl_cbc}.  Both exclusions
are shown in terms of redshift, where we assume a flat $\Lambda$CDM
cosmology with Hubble constant $H_0 = 70\,\mathrm{km\,s^{-1}Mpc^{-1}}$, 
dark matter content $\Omega_\text{M} = 0.27$ and dark energy content 
$\Omega_\Lambda = 0.73$ \citep{Komatsu10}.  The exclusion at
low redshift is dictated by the number of analyzed GRBs and at high
redshift by the typical sensitive range of the search.  
These exclusions assume 100\% purity of the GRB sample.  For purity $p$ 
the cumulative distribution should be rescaled by $1/p$; for instance, 
only one third of our short GRB sample has a $T_{90} < 2\,\mathrm{s}$. 
For comparison, each figure also shows the distribution of measured 
GRB redshifts, for all {\it Swift} GRBs (Fig.~\ref{fig:zExcl_csg150}) 
or for all short GRBs (Fig.~\ref{fig:zExcl_cbc}).  While the  
distribution of GRBs with measured redshifts includes various 
observational biases compared to the distribution of all GRBs  
detected electromagnetically (and on which we perform exclusions), 
it is clear that the exclusions from the current coalescence
and GW burst searches are not sufficient to put any additional constraint
on the nature of GRBs.

While this search for gravitational wave signals in coincidence with
observed GRBs was not at the sensitivity necessary to detect such
coincidences, it is interesting to consider the chances of detection
with the Advanced LIGO/Virgo detectors \citep{aVirgo,aLIGO}, which should become operational
in 2015. At their design sensitivity, these detectors should offer a factor
of 10 improvement in distance sensitivity to both GW burst and coalescence signals, 
dramatically improving the chances to make a
gravitational-wave observation of an electromagnetically detected GRB.

In Fig.~\ref{fig:zExcl_csg150} and Fig.~\ref{fig:zExcl_cbc} we extrapolate the
current exclusion curves to the advanced detector era, by assuming a factor
10 increase in sensitivity of the GW detectors and a factor 5 increase in the 
number of GRBs analyzed (equivalent to approximately 2.5 years of
live observing time at the rate that GRBs are currently being reported).
These extrapolations show that detection is quite possible in the advanced
detector era. Even if a detection is not made, targeted gravitational wave
searches will allow us to place
astrophysically relevant constraints on GRB population models.

For long GRBs, the Advanced LIGO/Virgo detectors will be able to test
optimistic scenarios for GW emission -- those that produce
$\sim 10^{-2}\,\mathrm{M_\odot c^2}$ in the most sensitive frequency band
of the detectors. The sensitive range for these systems will include
the local population of sub-luminous GRBs that produce the
low-redshift excess in Fig.~\ref{fig:zExcl_csg150}.  We note, however,
that GW burst emission with significantly lower $E_\text{GW}$ or at 
non-optimal frequencies is unlikely to be detectable. 

For short GRBs, a coincident GW detection appears quite possible. 
This conclusion is consistent with simple estimates such as 
that of \citet{Metzger11}, who estimate a coincident 
observation rate of $3\,\mathrm{yr^{-1}}$ ($0.3\,\mathrm{yr^{-1}}$) 
for NS--BH systems (NS--NS systems) with the advanced detectors.
The precise rate of occurrence will depend on the typical masses of the
compact objects; we are sensitive to NS--BH systems at a larger distance than
NS--NS systems. 
The distribution of binary component spins and the jet opening angle 
will also affect the received GW signal strength.  
The detection rate will also depend on the shape of the short
GRB cumulative distance curve at low redshift. 
One must also remember that we used
a very optimistic definition of short GRBs to avoid missing a potential signal.
It is likely that some of the short GRBs
that we analyzed for coalescence signals were not produced by a
compact binary progenitor.
Even in the case that no coalescence signals are detected in coincidence with short GRBs
in the advanced-detector era, it should be possible to place astrophysically
interesting constraints on the physical characteristics of progenitors
of short GRBs.

Finally, we note that these extrapolations carry a number of other 
uncertainties. In particular, the actual performance of future detectors is unknown. 
Furthermore, the extrapolations depend on how well the sky will be 
covered by gamma-ray satellites in 2015 and later compared to the present day.

\section{Conclusion}
\label{sec:conclusion}

We performed searches for gravitational waves coincident with gamma-ray bursts
during the 2009-2010 science runs of LIGO and Virgo.  In
total we analyzed \nAnalyzedGRB\ GRBs using two different analysis
methods.  A GW burst search looked for unmodeled transient signals, as
expected from massive stellar collapses, and a focused search looked
for coalescence signals from the merger of two compact objects, as expected
for short GRBs.  We did not detect any gravitational wave
in coincidence with a GRB in either search.  We set lower limits on the
distance of each GRB with the GW burst search, and of the short
GRBs with the coalescence search.  The median exclusion distances are
$17\,\mathrm{Mpc} \, (E_\text{GW}/10^{-2}\,\mathrm{M_\odot
  c^2})^{1/2}$ at 150\,Hz for the GW burst search and \distNSNS~Mpc
(\distNSBH~Mpc) for NS-NS (NS-BH) systems for the coalescence search, given
the priors on the source parameters described in
Sec.~\ref{sec:search}.  

These two searches are more sensitive than the corresponding
all-sky searches of the same data \citep{cbcLowMassS6,2012arXiv1202.2788T}, due to the
more focused analysis possible given the trigger time and sky
position information provided by the GRB satellites.  This improvement 
is as much as a factor of $\sim$2 in distance for the GW burst search.  
Additionally, our exclusion distances are greater because each source 
can be presumed to be favorably oriented relative to our line of sight, 
with limits on misalignment set by inferences of short and long GRB jet 
opening angles.  
Further theoretical studies of GRB central engines and observational
constraints on jet breaks and jet opening angles could allow this and
future studies to refine their constraints \emph{a posteriori}.
Additionally, improved methods of classification of GRBs, and in 
particular of identifying GRBs with possible binary progenitors with 
a lower false assignment rate, will improve the performance of our 
population estimates.

The LIGO and Virgo detectors are currently undergoing a major
upgrade, implementing new techniques to greatly increase their sensitivity, 
and are expected to begin operations by 2015. 
With these advanced detectors our chances to make a coincident GW
observation of a GRB are good, but depend strongly on the advanced detectors
running an extended science run at design sensitivity and the number of GRBs
that will be observed electromagnetically.
Therefore it is of utmost importance to have GRB satellites operating during the 
advanced detector era to provide electromagnetic triggers around which a more 
sensitive search for gravitational waves can be performed.

\acknowledgments
We are indebted to the observers of the electromagnetic events
and the Gamma-ray burst Coordinates Network for providing us with valuable data.
The authors gratefully acknowledge the support of the United States
National Science Foundation for the construction and operation of the
LIGO Laboratory, the Science and Technology Facilities Council of the
United Kingdom, the Max-Planck-Society, and the State of
Niedersachsen/Germany for support of the construction and operation of
the GEO600 detector, and the Italian Istituto Nazionale di Fisica
Nucleare and the French Centre National de la Recherche Scientifique
for the construction and operation of the Virgo detector. The authors
also gratefully acknowledge the support of the research by these
agencies and by the Australian Research Council, 
the International Science Linkages program of the Commonwealth of Australia,
the Council of Scientific and Industrial Research of India, 
the Istituto Nazionale di Fisica Nucleare of Italy, 
the Spanish Ministerio de Econom\'ia y Competitividad,
the Conselleria d'Economia Hisenda i Innovaci\'o of the
Govern de les Illes Balears, the Foundation for Fundamental Research
on Matter supported by the Netherlands Organisation for Scientific Research, 
the Polish Ministry of Science and Higher Education, the FOCUS
Programme of Foundation for Polish Science,
the Royal Society, the Scottish Funding Council, the
Scottish Universities Physics Alliance, The National Aeronautics and
Space Administration, the Carnegie Trust, the Leverhulme Trust, the
David and Lucile Packard Foundation, the Research Corporation, and
the Alfred P. Sloan Foundation.
This document has been assigned LIGO Laboratory document
number LIGO-P1000121-v\dccversion.

\appendix

\section{Weighted binomial test}
\label{sec:weightedBinomial}

In a search for GWs associated with GRBs, data corresponding to each
GRB are analyzed independently. The results of these independent
analyses need to be combined into a single GW (non-)detection
statement, which accounts for both the possibility of a single loud GW
event or a population of weak GW signals.  This weighted binomial test
is an extension of the binomial test used to look for an excess of
weak gravitational wave signatures in previous searches for GW bursts
associated with GRBs~\citep{burstGrbS234,burstGrbS5}.

The binomial test considers the set
$\{p_i\}_{1 \leq i \leq  N_\text{GRB}}$ 
of p-values obtained for a population of
$N_\text{GRB}$ analyzed GRBs, sorted increasingly. The smallest
$N_\text{tail} = 0.05 N_\text{GRB}$ of these p-values are used to search
for an excess of weak signals.
The binomial probability, under the null hypothesis, of obtaining at least $k$
events with p-values less than the actual $k$-th p-value $p_k$ is calculated for
$1 \leq k \leq N_\text{tail} $ and the minimum of these probabilities is used
as a detection statistic:
\begin{equation}
S_\text{binomial} =  -\log \min_{1\leq k \leq N_\text{tail}} \sum_{l\geq k} \binom{N}{l} p_k^l
  (1-p_k)^{N-l} \, .
\end{equation}
$S_\text{binomial}$ looks for a deviation of the p-value distribution when compared to the
uniform distribution expected from background, in the low p-value
region where an excess of weak gravitational wave signals might be
observable. However, this detection statistic does not take into account the
relative \emph{a priori} GW detection probabilities; that is the sensitive
volumes of the GW search associated with each GRB trigger, which
depends on the GRB sky position and the performance of GW detectors at
that time. To reduce the contribution of GRBs for which the GW
detector sensitivity is poor we construct a weighted binomial test
\citep{thesisWas} as
follows:
\begin{enumerate}
\item Based on the background and sensitivity to simulated signals, compute the
  distance $d_k(i)$ at which the detection efficiency is equal to 50\%
  for GRB $k$ and signal emission model $i$.
\item Compute the relative volume ratio $R_k(i) = d_k(i)^3/\max_l
  d_l(i)^3$ for model $i$ compared to the most sensitive GRB.
\item Average the relative volume ratio over the different models
  $R_k = \textrm{mean}_{i} R_k(i)$.
\item Sort the penalized p-values $p_k/R_k$ in increasing
  order, and compute the detection statistic
  \begin{equation}
    S_\text{weighted} = -\log \min_{1\leq k \leq N_\text{tail}} 
  \binom{N}{k}\prod_{l \leq k}\frac{p_l}{R_l} \, .
  \end{equation}
\end{enumerate}
For the GW burst search we use the 2 coalescence models and 3 GW burst
models given in Secs.~\ref{sec:burst} and~\ref{sec:cbc} to construct
the weighted binomial test, in order to include a range of possible emission
models. For the coalescence search we use only the 2 coalescence models,
which is appropriate for that more focused modeled search.

\section{Population exclusion method}
\label{sec:appendix:popExclMethod}

A lack of detection can be interpreted individually for each
analyzed GRB with an exclusion distance for given GW emission models.
But the set of analyzed GRBs can also be considered as a whole, to derive
constraints on the population of GRBs detected by $\gamma$-ray
satellites.  To perform such an exclusion we use 
a simple population model with all GRB progenitors having the same GW
emission (standard sirens), and with a distance distribution with two
components: a fraction $F$ of GRBs distributed with a
constant co-moving rate density up to a luminosity distance $R$, and a
fraction $1-F$ at effectively infinite distance. This simple model
yields a parameterization of astrophysical GRB distance distribution
models that predict a uniform local rate density and a more complex
dependence at redshift $>0.1$, as the large-redshift part of the
distribution is well beyond the sensitivity of current GW detectors.

For this population model we set a frequentist limit on the $F$ and
$R$ parameters by excluding all $(F,R)$ which have a 90\% or greater
chance of yielding an event with ranking statistic greater than the
largest value actually measured for any of the analyzed GRBs. In our
computations we  assume a flat
$\Lambda$CDM cosmology with Hubble constant $H_0 =
70\,\mathrm{km\,s^{-1}Mpc^{-1}}$, dark matter content $\Omega_\text{M}
= 0.27$ and dark energy content $\Omega_\Lambda = 0.73$
\citep{Komatsu10}.

In practice for each GRB $k$ we measure the efficiency $e_k(r)$ as a
function of luminosity distance $r$, for a given GW source model of
yielding an event with ranking statistic greater than the largest
value actually measured. This efficiency is integrated over the volume
of radius $R$, where the sources are distributed with constant
rate-density. Using the volume element for a flat cosmology, 
\begin{equation}
  \frac{\mathrm{d}V}{\mathrm{d}r} = \frac{4 \pi
    r^2}{ (1+z) \left[(1+z)^2 +  \frac{r H_0}{c} \sqrt{ \Omega_\text{M}(1+z)^3 + \Omega_\Lambda}  
     \right]} \, ,
\end{equation}
we integrate the efficiency as a function of luminosity
distance over the considered volume
\begin{equation}
  E_k(R) = \frac{\int_0^R e_k(r)   \frac{\mathrm{d}V}{\mathrm{d}r} \frac{\mathrm{d}
      r}{1 + z}}{\int_0^R \frac{\mathrm{d}V}{\mathrm{d}r} \frac{\mathrm{d}
      r}{1 + z}} \, ,
\end{equation}
where the additional $1/(1+z)$ factor accounts for the redshift of the rate.
This volume efficiency is the probability for a GRB progenitor 
to yield an event with higher ranking statistic than the value
actually measured, under the assumption that the GRB has a distance
distributed uniformly within the volume of radius $R$.
This can then be extended to a subset of GRBs
$\{k_1,\ldots,k_M\}$ all within the local volume of radius $R$, to
construct the probability of at least one of them yielding a higher
ranking statistic than the measured one:
\begin{equation}
  E_{\{k_1,\ldots,k_M\}}(R) = 1 - \!\! \prod_{k \in \{k_1,\ldots,k_M\}} \!\! \left[1 - E_k(R)\right] \, .
\end{equation}

However, our model predicts that a fraction of GRBs $1-F$ will
originate from distances larger than $R$, and thus be unobservable.
For a given fraction $F$, the distribution of the number $J$ of GRBs
in the local volume for a sample of $N$ GRBs is binomial, and all 
subsets $\{k_1,\ldots,k_J\}$ of $\lsem 1, N\rsem$ have equal
probability, given that we assume no knowledge of
\textit{which} of the GRBs are in the local volume and which are not.
The probability of there being exactly $J$ GRBs in the local volume is
given by the binomial probability,
\begin{equation}
  p(J | N) =  \binom{N}{J} F^J (1-F)^{N-J} \, ,
\end{equation}
and thus the probability of having a given subset of GRBs within $R$ is 
\begin{equation}
  p(\{k_1,\ldots,k_J\}) = F^J (1-F)^{N-J} \, .
\end{equation}

We can then obtain the probability that we would have observed a
gravitational-wave signal with higher ranking statistic than the one 
actually measured for at least one of the GRBs, as a function of
$F$ and $R$, by summing over the probability of all possible
configurations. This is given by
\begin{equation}
  E_F(R) = \sum_{J=0}^N \sum_{\{k_1, \ldots\, k_J\} \subset \lsem
    1,\,N\rsem} F^J (1-F)^{N-J}  E_{\{k_1,\ldots,k_J\}}(R) \, ,
\end{equation}
and parameters $(F,R)$ for which $E_F(R) > 0.9$ are excluded at 90\%
confidence. That is, we exclude any cumulative distance distribution
model that passes through an excluded $(F,R)$ point and which is 
uniform up to that point.

This framework can also be expanded to include a mixed sample of GRBs,
with a fraction $p$ of GRBs following the given standard siren model,
and a fraction $1-p$ without any significant GW emission. In that case
the cumulative distance distribution of the GRBs following the
standard siren model is excluded whenever $E_{pF}(R) > 0.9$; that is, 
the exclusion curve is scaled by a $1/p$ factor compared to the pure
sample case.

\section{Result tables}

\begin{longtable*}{llrrlrrrrr}
  \caption{\textsc{Short GRB sample and search results}\\ 
  \label{tab:shortGRB}}
  \\\hline
 & $\vphantom{{A^A}^A}$ &  &  &  &   
 \multicolumn{4}{c}{Exclusion (Mpc)}\\
& UTC& & & network \& & \multicolumn{2}{c}{GW burst at} & & \\
GRB name & time & RA & Dec &  time window & 150\,Hz&
300\,Hz& NS-NS & NS-BH & $\gamma$-ray detector\\
\hline
\endfirsthead

\multicolumn{10}{c}{ \tablename\ \thetable{} \emph{continued} }\\ \\[-1em]\hline
 & $\vphantom{{A^A}^A}$ &  &  &  & 
 \multicolumn{4}{c}{Exclusion (Mpc)}\\
& UTC & & & network \& & \multicolumn{2}{c}{GW burst at} & & \\
GRB name & time & RA & Dec &  time window & 150\,Hz&
300\,Hz & NS-NS & NS-BH& $\gamma$-ray detector\\
\hline
\endhead

  \hline
  \endfoot

  \hline
\multicolumn{10}{c}{\begin{minipage}{0.89\linewidth}
$\vphantom{{A^A}^A}$ 
Information and limits on associated GW emission for each of the
analyzed GRBs that were classified by us as short.  The first four columns
are: the GRB name in YYMMDD format 
or the Fermi/GBM trigger ID for GBM triggers classified as a GRB without an 
available GRB name (see \href{http://heasarc.gsfc.nasa.gov/W3Browse/fermi/fermigbrst.html}{http://heasarc.gsfc.nasa.gov/W3Browse/fermi/fermigbrst.html} and 
\citet{Paciesas12}); the trigger time (numbers in
parentheses denote the time in seconds by which the trigger was shifted
for the coalescence search following visual inspection of the lightcurve); 
and the sky position used for the GW search (right ascension and declination). 
Both a $^\sharp$ and a $^\ddagger$ indicate that, although the formal duration
of this GRB is longer than 4 s ($^\ddagger$), or unavailable ($^\sharp$), the 
GRB was analyzed as a short GRB because of a prominent short spike at the 
beginning of the lightcurve (see Sec.~\ref{sec:grbsample}). 
The fifth column gives the gravitational wave detector network used; a
$^*$ indicates when the shorter on-source window starting 120\,s before the
trigger is used for the GW burst search, and a $^\dagger$ when the on-source window 
is extended to cover the GRB duration ($T_{90} > 60\,\mathrm{s}$). 
A $^\Diamond$ indicates the use of only H1L1 data for the burst search, 
because of data quality requirements.
Columns 6-9 display the result of the search: the 90\%
confidence lower limits on the distance to the GRB for different
waveform models.  A standard siren energy emission of
$E_\text{GW} = 10^{-2}\,\mathrm{M_\odot c^2}$ is assumed for the
circular sine-Gaussian GW burst models; these limits are not
available for 4 short GRBs which were not analyzed by GW burst search. 
The last column gives the 
$\gamma$-ray detector that provided the sky location used for the search.
For GRB 090802, IPN triangulation from Konus-WIND, INTEGRAL and \emph{Fermi} was 
used to further constrain the sky position.   
The intersection of the IPN and \emph{Fermi} error regions 
was used to place search points using the method described in
\citet{Predoi:2011aa}. For this GRB, the quoted right ascension and declination
corresponds to the centre of the \emph{Fermi} error region. For
IPN
localizations a complete list of detectors can be found on the
project trigger page,
\href{http://www.ssl.berkeley.edu/ipn3/masterli.txt}{http://www.ssl.berkeley.edu/ipn3/masterli.txt}.
\end{minipage}}
  \endlastfoot
090720B$^\ddagger$ & 17:02:56 & $ 13^{\mathrm{h}} 31^{\mathrm{m}} 59^{\mathrm{s}}$ & $ -54^{\circ} 48' $ & L1V1 & 7.1 & 3.8 & 8.6 & 16.0 & GBM\\
090802A & 05:39:03 & $ 5^{\mathrm{h}} 37^{\mathrm{m}} 19^{\mathrm{s}}$ & $34^{\circ} 05' $ & H1L1V1 & 7.3 & 2.6 & 6.5 & 11.3 & GBM and IPN\\
090815C & 23:21:39 & $ 4^{\mathrm{h}} 17^{\mathrm{m}} 57^{\mathrm{s}}$ & $ -65^{\circ} 57' $ & H1L1V1$^*$ & 29.8 & 12.0 & 24.6 & 44.3 & BAT\\
090820B$^\ddagger$ & 12:13:16 & $ 21^{\mathrm{h}} 13^{\mathrm{m}} 02^{\mathrm{s}}$ & $ -18^{\circ} 35' $ & H1V1 & 12.2 & 5.8 & 15.1 & 26.3 & GBM\\
090831A$^\ddagger$ & 07:36:36 & $ 9^{\mathrm{h}} 40^{\mathrm{m}} 23^{\mathrm{s}}$ & $ 50^{\circ} 58' $ & H1V1$^\dagger$ & 7.2 & 2.1 & 4.6 & 8.9 & GBM\\
090927 & 10:07:16$\,$($+$1) & $ 22^{\mathrm{h}} 55^{\mathrm{m}} 42^{\mathrm{s}}$ & $ -70^{\circ} 58' $ & H1L1V1 & 16.0 & 9.0 & 19.8 & 35.1 & BAT\\
091018$^\ddagger$ & 20:48:19 & $ 2^{\mathrm{h}} 08^{\mathrm{m}} 46^{\mathrm{s}}$ & $ -57^{\circ} 33' $ & H1V1 & $-$ & $-$ & 5.2 & 10.0 & BAT\\
091126A & 07:59:24 & $ 5^{\mathrm{h}} 33^{\mathrm{m}} 00^{\mathrm{s}}$ & $ -19^{\circ} 16'$ & H1V1 & $-$ & $-$ & 13.9 & 25.3 & GBM\\
091127$^\ddagger$ & 23:25:45 & $ 2^{\mathrm{h}} 26^{\mathrm{m}} 19^{\mathrm{s}}$ & $ -18^{\circ} 57' $ & L1V1$^*$ & 5.9 & 2.3 & 3.1 & 4.9 & BAT\\
091208B$^\ddagger$ & 09:49:57 & $ 1^{\mathrm{h}} 57^{\mathrm{m}} 39^{\mathrm{s}}$ & $ 16^{\circ} 53' $ & H1V1 & $-$ & $-$ & 11.4 & 20.6 & BAT\\
100111A$^\ddagger$ & 04:12:49 & $ 16^{\mathrm{h}} 28^{\mathrm{m}} 06^{\mathrm{s}}$ & $ 15^{\circ} 32' $ & H1L1 & 18.8 & 8.6 & 17.7 & 30.4 & BAT\\
100206A & 13:30:05 & $ 3^{\mathrm{h}} 08^{\mathrm{m}} 40^{\mathrm{s}}$ & $ 13^{\circ} 10' $ & H1L1 & 21.0 & 8.8 & 19.1 & 34.1 & BAT\\
100213A & 22:27:48 & $ 23^{\mathrm{h}} 17^{\mathrm{m}} 30^{\mathrm{s}}$ & $ 43^{\circ} 22' $ & H1L1 & 22.4 & 10.0 & 24.5 & 46.3 & BAT\\
100216A & 10:07:00 & $ 10^{\mathrm{h}} 17^{\mathrm{m}} 03^{\mathrm{s}}$ & $ 35^{\circ} 31' $ & H1L1 & 29.1 & 13.0 & 22.7 & 40.1 & BAT\\
100316B & 08:01:36 & $ 10^{\mathrm{h}} 54^{\mathrm{m}} 00^{\mathrm{s}}$ & $-45^{\circ} 28'$ & H1L1 & $-$ & $-$ & 2.1 & 3.7 & BAT\\
100322B & 07:06:18 & $ 5^{\mathrm{h}} 05^{\mathrm{m}} 57^{\mathrm{s}}$ & $ 42^{\circ} 41' $ & H1L1 & 18.5 & 7.4 & 14.8 & 25.4 & BAT\\
100325B$^\ddagger$ & 05:54:43 & $ 13^{\mathrm{h}} 56^{\mathrm{m}} 33^{\mathrm{s}}$ & $ -79^{\circ} 06' $ & H1L1 & 21.8 & 8.7 & 19.0 & 34.3 & GBM\\
100328A & 03:22:44 & $ 10^{\mathrm{h}} 23^{\mathrm{m}} 45^{\mathrm{s}}$ & $ 47^{\circ} 02' $ & H1L1 & 28.9 & 12.4 & 30.1 & 51.3 & GBM\\
100515A$^\ddagger$ & 11:13:09 & $ 18^{\mathrm{h}} 21^{\mathrm{m}} 52^{\mathrm{s}}$ & $ 27^{\circ} 01' $ & H1L1 & 38.2 & 17.1 & 37.1 & 64.5 & GBM\\
100517D$^\ddagger$ & 03:42:08 & $ 16^{\mathrm{h}} 14^{\mathrm{m}} 21^{\mathrm{s}}$ & $ -10^{\circ} 22' $ & H1L1 & 3.4 & 2.7 & 7.7 & 12.1 & GBM\\
100628A & 08:16:40 & $ 15^{\mathrm{h}} 03^{\mathrm{m}} 46^{\mathrm{s}}$ & $ -31^{\circ} 39' $ & H1L1 & 20.6 & 8.3 & 20.7 & 36.7 & BAT\\
100717446$^\sharp$ & 10:41:47 & $ 20^{\mathrm{h}} 17^{\mathrm{m}} 14^{\mathrm{s}}$ & $ 19^{\circ} 32' $ & H1L1 & 31.3 & 13.2 & 26.5 & 46.1 & GBM\\
100816A & 00:37:51 & $ 23^{\mathrm{h}} 26^{\mathrm{m}} 57^{\mathrm{s}}$ & $ 26^{\circ} 34' $ & L1V1 & 9.5 & 5.8 & 6.6 & 11.5 & BAT\\
100905A & 15:08:14$\,$($-$1) & $ 2^{\mathrm{h}} 06^{\mathrm{m}} 10^{\mathrm{s}}$ & $ 14^{\circ} 55' $ & H1L1V1$^\Diamond$ & 17.3 & 6.3 & 11.5 & 19.6 & BAT\\
100924A$^\ddagger$ & 03:58:08 & $ 0^{\mathrm{h}} 02^{\mathrm{m}} 41^{\mathrm{s}}$ & $ 7^{\circ} 00' $ & H1L1V1$^{\dagger\,\Diamond}$ & 29.2 & 12.0 & 22.8 & 39.4 & BAT\\
100928A & 02:19:52$\,$($+$1) & $ 14^{\mathrm{h}} 52^{\mathrm{m}} 08^{\mathrm{s}}$ & $ -28^{\circ} 33' $ & H1L1V1$^\Diamond$ & 26.1 & 10.1 & 20.1 & 35.1 & BAT\\
  \hline
\end{longtable*}

\begin{longtable*}{lcrrlrrr}
\caption{\textsc{Long GRB sample and search results} \label{tab:longGRB}}  \\
 \\[-0.75em]\hline
 & $\vphantom{{A^A}^A}$ &  &  &  & 
 \multicolumn{2}{c}{Exclusion (Mpc)}\\
& UTC & & & network \& &\multicolumn{2}{c}{GW burst at} & \\
GRB name & time & RA & Dec &  time window & 150\,Hz &
300\,Hz & $\gamma$-ray detector\\
\hline
\endfirsthead

\multicolumn{8}{c}{ \tablename\ \thetable{} \emph{continued} }\\ \\\hline
 & $\vphantom{{A^A}^A}$ &  &  &  & 
 \multicolumn{2}{c}{Exclusion (Mpc)}\\
& UTC & & & network \& &\multicolumn{2}{c}{GW burst at} & \\
GRB name & time & RA & Dec &  time window & 150\,Hz &
300\,Hz & $\gamma$-ray detector\\
\hline
\endhead

\hline
\endfoot

\hline \multicolumn{8}{c}{\begin{minipage}{0.71\linewidth} 
    $\vphantom{{A^A}^A}$ 
    Information and limits on associated GW emission for each of the
    analyzed GRBs that were classified as long.  The first four columns
    are: the GRB name in YYMMDD format or the Fermi/GBM trigger ID for GBM
    triggers classified as a GRB without an available GRB name (see 
    \href{http://heasarc.gsfc.nasa.gov/W3Browse/fermi/fermigbrst.html}{http://heasarc.gsfc.nasa.gov/W3Browse/fermi/fermigbrst.html} 
    and \citet{Paciesas12}); the trigger time; and the sky position used 
    for the GW search (right ascension and declination).   
    The fifth column gives the gravitational wave detector network used; a
    $^*$ indicates when the shorter on-source window starting 120\,s before the
    trigger is used, and a $^\dagger$ when the on-source window 
    is extended to cover the GRB duration ($T_{90} > 60\,\mathrm{s}$). 
    Columns 6-7 display the result of the search: the 90\%
    confidence lower limits on the distance to the
    GRB for the circular sine-Gaussian GW burst models at 150\,Hz and
    300\,Hz.  A standard siren energy emission of
    $E_\text{GW} = 10^{-2}\,\mathrm{M_\odot c^2}$ is assumed.  The last 
    column gives the $\gamma$-ray detector that provided the 
    sky location used for the search. For IPN
    localizations a complete list of detectors can be found on the 	    
    project trigger page, 
    \href{http://www.ssl.berkeley.edu/ipn3/masterli.txt}{http://www.ssl.berkeley.edu/ipn3/masterli.txt}.
\end{minipage}}
\endlastfoot
090709B & 15:07:42 & $ 6^{\mathrm{h}} 14^{\mathrm{m}} 05^{\mathrm{s}}$ & $ 64^{\circ} 05' $ & L1V1 & 12.4 & 6.1 & BAT\\
090717A & 00:49:32 & $ 5^{\mathrm{h}} 47^{\mathrm{m}} 19^{\mathrm{s}}$ & $ -64^{\circ} 11' $ & H1V1$^*$$^\dagger$ & 19.9 & 9.7 & GBM\\
090719 & 01:31:26 & $ 22^{\mathrm{h}} 45^{\mathrm{m}} 04^{\mathrm{s}}$ & $ -67^{\circ} 52' $ & H1V1 & 10.6 & 6.3 & GBM\\
090720A & 06:38:08 & $ 13^{\mathrm{h}} 34^{\mathrm{m}} 46^{\mathrm{s}}$ & $ -10^{\circ} 20' $ & L1V1 & 12.5 & 6.4 & BAT\\
090726B & 05:14:07 & $ 16^{\mathrm{h}} 01^{\mathrm{m}} 48^{\mathrm{s}}$ & $ 36^{\circ} 45' $ & H1L1V1 & 20.2 & 6.4 & GBM\\
090726 & 22:42:27 & $ 16^{\mathrm{h}} 34^{\mathrm{m}} 43^{\mathrm{s}}$ & $ 72^{\circ} 52' $ & H1V1$^\dagger$ & 17.1 & 9.3 & BAT\\
090727 & 22:42:18 & $ 21^{\mathrm{h}} 03^{\mathrm{m}} 40^{\mathrm{s}}$ & $ 64^{\circ} 56' $ & L1V1$^\dagger$ & 10.4 & 4.9 & BAT\\
090727B & 23:32:29 & $ 22^{\mathrm{h}} 53^{\mathrm{m}} 25^{\mathrm{s}}$ & $ -46^{\circ} 42' $ & L1V1 & 3.3 & 1.8 & IPN\\
090802B & 15:58:23 & $ 17^{\mathrm{h}} 48^{\mathrm{m}} 04^{\mathrm{s}}$ & $ -71^{\circ} 46' $ & H1L1V1 & 20.5 & 8.3 & GBM\\
090807 & 15:00:27 & $ 18^{\mathrm{h}} 14^{\mathrm{m}} 57^{\mathrm{s}}$ & $ 10^{\circ} 17' $ & H1V1$^\dagger$ & 9.8 & 5.3 & BAT\\
090809 & 17:31:14 & $ 21^{\mathrm{h}} 54^{\mathrm{m}} 39^{\mathrm{s}}$ & $ -0^{\circ} 05' $ & H1L1V1 & 19.2 & 6.4 & BAT\\
090809B & 23:28:14 & $ 6^{\mathrm{h}} 20^{\mathrm{m}} 60^{\mathrm{s}}$ & $ 0^{\circ} 10' $ & L1V1 & 9.5 & 4.9 & GBM\\
090810A & 15:49:07 & $ 11^{\mathrm{h}} 15^{\mathrm{m}} 43^{\mathrm{s}}$ & $ -76^{\circ} 24' $ & H1V1 & 14.6 & 7.1 & GBM\\
090814A & 00:52:19 & $ 15^{\mathrm{h}} 58^{\mathrm{m}} 27^{\mathrm{s}}$ & $ 25^{\circ} 35' $ & L1V1$^\dagger$ & 9.9 & 6.1 & BAT\\
090814B & 01:21:01 & $ 4^{\mathrm{h}} 19^{\mathrm{m}} 05^{\mathrm{s}}$ & $ 60^{\circ} 35' $ & L1V1 & 10.9 & 5.6 & IBIS\\
090814D & 22:47:28 & $ 20^{\mathrm{h}} 30^{\mathrm{m}} 35^{\mathrm{s}}$ & $ 45^{\circ} 43' $ & H1L1V1 & 17.2 & 6.3 & GBM\\
090815A & 07:12:12 & $ 2^{\mathrm{h}} 44^{\mathrm{m}} 07^{\mathrm{s}}$ & $ -2^{\circ} 44' $ & H1L1V1$^\dagger$ & 5.4 & 1.3 & GBM\\
090815B & 10:30:41 & $ 1^{\mathrm{h}} 25^{\mathrm{m}} 40^{\mathrm{s}}$ & $ 53^{\circ} 26' $ & H1V1 & 10.6 & 5.6 & GBM\\
090815D & 22:41:46 & $ 16^{\mathrm{h}} 45^{\mathrm{m}} 02^{\mathrm{s}}$ & $ 52^{\circ} 56' $ & L1V1 & 15.7 & 6.9 & GBM\\
090823B & 03:10:53 & $ 3^{\mathrm{h}} 18^{\mathrm{m}} 07^{\mathrm{s}}$ & $ -17^{\circ} 35' $ & L1V1 & 5.9 & 2.7 & GBM\\
090824A & 22:02:19 & $ 3^{\mathrm{h}} 06^{\mathrm{m}} 35^{\mathrm{s}}$ & $ 59^{\circ} 49' $ & H1V1 & 9.4 & 4.8 & GBM\\
090826 & 01:37:31 & $ 9^{\mathrm{h}} 22^{\mathrm{m}} 28^{\mathrm{s}}$ & $ -0^{\circ} 07' $ & H1V1 & 2.7 & 0.4 & GBM\\
090827 & 19:06:26 & $ 1^{\mathrm{h}} 13^{\mathrm{m}} 44^{\mathrm{s}}$ & $ -50^{\circ} 54' $ & H1V1 & 15.1 & 8.6 & BAT\\
090829B & 16:50:40 & $ 23^{\mathrm{h}} 39^{\mathrm{m}} 57^{\mathrm{s}}$ & $ -9^{\circ} 22' $ & H1V1$^\dagger$ & 9.0 & 4.8 & GBM\\
090926B & 21:55:48 & $ 3^{\mathrm{h}} 05^{\mathrm{m}} 14^{\mathrm{s}}$ & $ -38^{\circ} 60' $ & H1L1V1$^\dagger$ & 19.1 & 7.5 & BAT\\
090929A & 04:33:03 & $ 3^{\mathrm{h}} 26^{\mathrm{m}} 47^{\mathrm{s}}$ & $ -7^{\circ} 20' $ & H1V1 & 8.7 & 4.4 & GBM\\
091003 & 04:35:45 & $ 16^{\mathrm{h}} 45^{\mathrm{m}} 33^{\mathrm{s}}$ & $ 36^{\circ} 35' $ & L1V1 & 8.8 & 3.2 & LAT\\
091017A & 20:40:24 & $ 14^{\mathrm{h}} 03^{\mathrm{m}} 11^{\mathrm{s}}$ & $ 25^{\circ} 29' $ & H1V1 & 7.5 & 4.9 & GBM\\
091018B & 22:58:20 & $ 21^{\mathrm{h}} 27^{\mathrm{m}} 19^{\mathrm{s}}$ & $ -23^{\circ} 05' $ & L1V1 & 3.6 & 1.6 & GBM\\
091019A & 18:00:40 & $ 15^{\mathrm{h}} 04^{\mathrm{m}} 07^{\mathrm{s}}$ & $ 80^{\circ} 20' $ & H1L1V1 & 20.4 & 8.2 & GBM\\
091020 & 21:36:44 & $ 11^{\mathrm{h}} 42^{\mathrm{m}} 54^{\mathrm{s}}$ & $ 50^{\circ} 59' $ & L1V1 & 9.5 & 5.2 & BAT\\
091026B & 11:38:48 & $ 9^{\mathrm{h}} 08^{\mathrm{m}} 19^{\mathrm{s}}$ & $ -23^{\circ} 39' $ & H1V1 & 5.1 & 3.0 & GBM\\
091030A & 19:52:26 & $ 2^{\mathrm{h}} 46^{\mathrm{m}} 40^{\mathrm{s}}$ & $ 21^{\circ} 32' $ & H1V1$^\dagger$ & 11.5 & 4.9 & GBM\\
091031 & 12:00:28 & $ 4^{\mathrm{h}} 46^{\mathrm{m}} 47^{\mathrm{s}}$ & $ -57^{\circ} 30' $ & H1V1$^\dagger$ & 14.2 & 6.2 & LAT\\
091103A & 21:53:51 & $ 11^{\mathrm{h}} 22^{\mathrm{m}} 24^{\mathrm{s}}$ & $ 11^{\circ} 18' $ & L1V1 & 5.1 & 2.2 & GBM\\
091109A & 04:57:43 & $ 20^{\mathrm{h}} 37^{\mathrm{m}} 00^{\mathrm{s}}$ & $ -44^{\circ} 11' $ & H1V1 & 13.0 & 7.3 & BAT\\
091109B & 21:49:03 & $ 7^{\mathrm{h}} 31^{\mathrm{m}} 00^{\mathrm{s}}$ & $ -54^{\circ} 06' $ & H1V1 & 14.2 & 8.7 & BAT\\
091115A & 04:14:50 & $ 20^{\mathrm{h}} 31^{\mathrm{m}} 02^{\mathrm{s}}$ & $ 71^{\circ} 28' $ & H1L1V1$^*$ & 14.1 & 7.2 & GBM\\
091122A & 03:54:20 & $ 7^{\mathrm{h}} 23^{\mathrm{m}} 26^{\mathrm{s}}$ & $ 0^{\circ} 34' $ & H1V1 & 11.9 & 5.5 & GBM\\
091123B & 01:55:59 & $ 22^{\mathrm{h}} 31^{\mathrm{m}} 16^{\mathrm{s}}$ & $ 13^{\circ} 21' $ & L1V1$^*$ & 9.8 & 5.4 & GBM\\
091128 & 06:50:34 & $ 8^{\mathrm{h}} 30^{\mathrm{m}} 45^{\mathrm{s}}$ & $ 1^{\circ} 44' $ & H1V1 & 6.7 & 3.1 & GBM\\
091202B & 01:44:06 & $ 17^{\mathrm{h}} 09^{\mathrm{m}} 59^{\mathrm{s}}$ & $ -1^{\circ} 54' $ & H1V1 & 10.5 & 4.6 & GBM\\
091202C & 05:15:42 & $ 0^{\mathrm{h}} 55^{\mathrm{m}} 26^{\mathrm{s}}$ & $ 9^{\circ} 05' $ & H1V1 & 12.6 & 6.2 & GBM\\
091202 & 23:10:04 & $ 9^{\mathrm{h}} 15^{\mathrm{m}} 18^{\mathrm{s}}$ & $ 62^{\circ} 33' $ & H1V1 & 14.4 & 6.5 & IBIS\\
091215A & 05:37:26 & $ 18^{\mathrm{h}} 52^{\mathrm{m}} 59^{\mathrm{s}}$ & $ 17^{\circ} 33' $ & H1L1$^*$ & 18.5 & 8.1 & GBM\\
091219A & 11:04:45 & $ 19^{\mathrm{h}} 37^{\mathrm{m}} 57^{\mathrm{s}}$ & $ 71^{\circ} 55' $ & H1L1V1 & 12.4 & 6.1 & GBM\\
091220A & 10:36:50 & $ 11^{\mathrm{h}} 07^{\mathrm{m}} 04^{\mathrm{s}}$ & $ 4^{\circ} 49' $ & H1L1V1 & 21.0 & 8.6 & GBM\\
091223B & 12:15:53 & $ 15^{\mathrm{h}} 25^{\mathrm{m}} 04^{\mathrm{s}}$ & $ 54^{\circ} 44' $ & H1L1 & 24.3 & 9.9 & GBM\\
091224A & 08:57:36 & $ 22^{\mathrm{h}} 04^{\mathrm{m}} 40^{\mathrm{s}}$ & $ 18^{\circ} 16' $ & H1L1 & 16.4 & 6.4 & GBM\\
091227A & 07:03:13 & $ 19^{\mathrm{h}} 47^{\mathrm{m}} 45^{\mathrm{s}}$ & $ 2^{\circ} 36' $ & H1L1V1 & 19.7 & 8.6 & GBM\\
100101A & 00:39:49 & $ 20^{\mathrm{h}} 29^{\mathrm{m}} 16^{\mathrm{s}}$ & $ -27^{\circ} 00' $ & H1L1$^*$ & 9.1 & 4.5 & GBM\\
100103A & 17:42:32 & $ 7^{\mathrm{h}} 29^{\mathrm{m}} 28^{\mathrm{s}}$ & $ -34^{\circ} 29' $ & H1L1V1 & 26.0 & 12.1 & IBIS\\
100112A & 10:01:17 & $ 16^{\mathrm{h}} 00^{\mathrm{m}} 33^{\mathrm{s}}$ & $ -75^{\circ} 06' $ & H1L1 & 15.7 & 7.5 & GBM\\
100201A & 14:06:17 & $ 8^{\mathrm{h}} 52^{\mathrm{m}} 24^{\mathrm{s}}$ & $ -37^{\circ} 17' $ & H1L1$^*$ & 25.9 & 10.8 & GBM\\
100212B & 13:11:45 & $ 8^{\mathrm{h}} 57^{\mathrm{m}} 04^{\mathrm{s}}$ & $ 32^{\circ} 13' $ & H1L1$^*$ & 6.8 & 3.7 & GBM\\
100213B & 22:58:34 & $ 8^{\mathrm{h}} 17^{\mathrm{m}} 16^{\mathrm{s}}$ & $ 43^{\circ} 28' $ & H1L1 & 15.6 & 7.1 & BAT\\
100219A & 15:15:46 & $ 10^{\mathrm{h}} 16^{\mathrm{m}} 48^{\mathrm{s}}$ & $ -12^{\circ} 33' $ & H1L1 & 17.3 & 6.7 & BAT\\
100221A & 08:50:26 & $ 1^{\mathrm{h}} 48^{\mathrm{m}} 28^{\mathrm{s}}$ & $ -17^{\circ} 25' $ & H1L1 & 29.1 & 12.5 & GBM\\
100225B & 05:59:05 & $ 23^{\mathrm{h}} 31^{\mathrm{m}} 24^{\mathrm{s}}$ & $ 15^{\circ} 02' $ & H1L1 & 20.0 & 8.3 & GBM\\
100225C & 13:55:31 & $ 20^{\mathrm{h}} 57^{\mathrm{m}} 04^{\mathrm{s}}$ & $ 0^{\circ} 13' $ & H1L1$^*$ & 13.7 & 5.8 & GBM\\
100228B & 20:57:47 & $ 7^{\mathrm{h}} 51^{\mathrm{m}} 57^{\mathrm{s}}$ & $ 18^{\circ} 38' $ & H1L1 & 13.8 & 6.8 & GBM\\
100301B & 05:21:46 & $ 13^{\mathrm{h}} 27^{\mathrm{m}} 24^{\mathrm{s}}$ & $ 19^{\circ} 50' $ & H1L1 & 24.0 & 9.1 & GBM\\
100315A & 08:39:12 & $ 13^{\mathrm{h}} 55^{\mathrm{m}} 35^{\mathrm{s}}$ & $ 30^{\circ} 08' $ & H1L1 & 43.5 & 16.9 & GBM\\
100316A & 02:23:00 & $ 16^{\mathrm{h}} 47^{\mathrm{m}} 48^{\mathrm{s}}$ & $ 71^{\circ} 49' $ & H1L1 & 18.2 & 6.9 & BAT\\
100316C & 08:57:59 & $ 2^{\mathrm{h}} 09^{\mathrm{m}} 14^{\mathrm{s}}$ & $ -67^{\circ} 60' $ & H1L1 & 39.4 & 16.5 & BAT\\
100324A & 00:21:27 & $ 6^{\mathrm{h}} 34^{\mathrm{m}} 26^{\mathrm{s}}$ & $ -9^{\circ} 44' $ & H1L1 & 34.3 & 12.6 & BAT\\
100324B & 04:07:36 & $ 2^{\mathrm{h}} 38^{\mathrm{m}} 41^{\mathrm{s}}$ & $ -19^{\circ} 17' $ & H1L1 & 20.2 & 7.3 & IPN\\
100325A & 06:36:08 & $ 22^{\mathrm{h}} 00^{\mathrm{m}} 57^{\mathrm{s}}$ & $ -26^{\circ} 28' $ & H1L1 & 36.7 & 14.1 & LAT\\
100326A & 07:03:05 & $ 8^{\mathrm{h}} 44^{\mathrm{m}} 57^{\mathrm{s}}$ & $ -28^{\circ} 11' $ & H1L1 & 13.1 & 5.6 & GBM\\
100331B & 21:08:38 & $ 20^{\mathrm{h}} 11^{\mathrm{m}} 56^{\mathrm{s}}$ & $ -11^{\circ} 04' $ & H1L1 & 14.1 & 6.1 & AGILE\\
100401A & 07:07:31 & $ 19^{\mathrm{h}} 23^{\mathrm{m}} 15^{\mathrm{s}}$ & $ -8^{\circ} 15' $ & H1L1$^\dagger$ & 17.6 & 6.2 & BAT\\
100410A & 08:31:57 & $ 8^{\mathrm{h}} 40^{\mathrm{m}} 04^{\mathrm{s}}$ & $ 21^{\circ} 29' $ & H1L1$^*$ & 3.6 & 1.5 & GBM\\
100410B & 17:45:46 & $ 21^{\mathrm{h}} 16^{\mathrm{m}} 59^{\mathrm{s}}$ & $ 37^{\circ} 26' $ & H1L1 & 28.3 & 12.7 & GBM\\
100418A & 21:10:08 & $ 17^{\mathrm{h}} 05^{\mathrm{m}} 25^{\mathrm{s}}$ & $ 11^{\circ} 27' $ & H1L1 & 26.5 & 11.8 & BAT\\
100420B & 00:12:06 & $ 8^{\mathrm{h}} 02^{\mathrm{m}} 11^{\mathrm{s}}$ & $ -5^{\circ} 49' $ & H1L1 & 32.6 & 12.5 & GBM\\
100420A & 05:22:42 & $ 19^{\mathrm{h}} 44^{\mathrm{m}} 21^{\mathrm{s}}$ & $ 55^{\circ} 45' $ & H1L1 & 19.1 & 7.5 & BAT\\
100423B & 05:51:25 & $ 7^{\mathrm{h}} 58^{\mathrm{m}} 40^{\mathrm{s}}$ & $ 5^{\circ} 47' $ & H1L1 & 18.6 & 6.4 & GBM\\
100425A & 02:50:45 & $ 19^{\mathrm{h}} 56^{\mathrm{m}} 38^{\mathrm{s}}$ & $ -26^{\circ} 28' $ & H1L1 & 41.4 & 15.6 & BAT\\
100427A & 08:31:55 & $ 5^{\mathrm{h}} 56^{\mathrm{m}} 41^{\mathrm{s}}$ & $ -3^{\circ} 28' $ & H1L1 & 25.5 & 11.2 & BAT\\
100502A & 08:33:02 & $ 8^{\mathrm{h}} 44^{\mathrm{m}} 02^{\mathrm{s}}$ & $ 18^{\circ} 23' $ & H1L1 & 4.0 & 2.6 & GBM\\
100507A & 13:51:15 & $ 0^{\mathrm{h}} 11^{\mathrm{m}} 36^{\mathrm{s}}$ & $ -79^{\circ} 01' $ & H1L1 & 23.9 & 7.8 & GBM\\
100508A & 09:20:42 & $ 5^{\mathrm{h}} 05^{\mathrm{m}} 03^{\mathrm{s}}$ & $ -20^{\circ} 45' $ & H1L1$^*$ & 47.3 & 18.2 & BAT\\
100516A & 08:50:41 & $ 18^{\mathrm{h}} 17^{\mathrm{m}} 38^{\mathrm{s}}$ & $ -8^{\circ} 12' $ & H1L1 & 28.9 & 11.1 & GBM\\
100516B & 09:30:38 & $ 19^{\mathrm{h}} 50^{\mathrm{m}} 43^{\mathrm{s}}$ & $ 18^{\circ} 40' $ & H1L1 & 35.0 & 14.3 & GBM\\
100517B & 01:43:08 & $ 6^{\mathrm{h}} 43^{\mathrm{m}} 43^{\mathrm{s}}$ & $ -28^{\circ} 59' $ & H1L1 & 18.6 & 7.0 & GBM\\
100517E & 05:49:52 & $ 0^{\mathrm{h}} 41^{\mathrm{m}} 45^{\mathrm{s}}$ & $ 4^{\circ} 26' $ & H1L1 & 26.8 & 10.5 & GBM\\
100517F & 15:19:58 & $ 3^{\mathrm{h}} 30^{\mathrm{m}} 55^{\mathrm{s}}$ & $ -71^{\circ} 52' $ & H1L1 & 23.9 & 10.2 & GBM\\
100517C & 03:09:50 & $ 2^{\mathrm{h}} 42^{\mathrm{m}} 31^{\mathrm{s}}$ & $ -44^{\circ} 19' $ & H1L1 & 35.9 & 13.6 & GBM\\
100526B & 19:00:38 & $ 0^{\mathrm{h}} 03^{\mathrm{m}} 06^{\mathrm{s}}$ & $ -37^{\circ} 55' $ & H1L1$^\dagger$ & 12.5 & 5.4 & BAT\\
100604A & 06:53:34 & $ 16^{\mathrm{h}} 33^{\mathrm{m}} 12^{\mathrm{s}}$ & $ -73^{\circ} 11' $ & H1L1 & 19.3 & 9.0 & GBM\\
100608A & 09:10:06 & $ 2^{\mathrm{h}} 02^{\mathrm{m}} 09^{\mathrm{s}}$ & $ 20^{\circ} 27' $ & H1L1 & 24.9 & 10.4 & GBM\\
100701B & 11:45:23 & $ 2^{\mathrm{h}} 52^{\mathrm{m}} 26^{\mathrm{s}}$ & $ -2^{\circ} 13' $ & H1L1 & 14.1 & 6.3 & GBM\\
100709A & 14:27:32 & $ 9^{\mathrm{h}} 30^{\mathrm{m}} 07^{\mathrm{s}}$ & $ 17^{\circ} 23' $ & H1L1 & 16.9 & 6.5 & GBM\\
100717372 & 08:55:06 & $ 19^{\mathrm{h}} 08^{\mathrm{m}} 14^{\mathrm{s}}$ & $ -0^{\circ} 40' $ & H1L1 & 27.3 & 10.7 & GBM\\
100719989 & 23:44:04 & $ 7^{\mathrm{h}} 33^{\mathrm{m}} 12^{\mathrm{s}}$ & $ 5^{\circ} 24' $ & H1L1 & 15.3 & 5.9 & GBM\\
100722291 & 06:58:24 & $ 2^{\mathrm{h}} 07^{\mathrm{m}} 14^{\mathrm{s}}$ & $ 56^{\circ} 14' $ & H1L1$^*$ & 17.7 & 6.7 & GBM\\
100725A & 07:12:52 & $ 11^{\mathrm{h}} 05^{\mathrm{m}} 52^{\mathrm{s}}$ & $ -26^{\circ} 40' $ & H1L1$^\dagger$ & 35.5 & 14.0 & BAT\\
100725B & 11:24:34 & $ 19^{\mathrm{h}} 20^{\mathrm{m}} 06^{\mathrm{s}}$ & $ 76^{\circ} 57' $ & H1L1$^\dagger$ & 25.9 & 10.8 & BAT\\
100727A & 05:42:17 & $ 10^{\mathrm{h}} 16^{\mathrm{m}} 44^{\mathrm{s}}$ & $ -21^{\circ} 25' $ & H1L1$^\dagger$ & 31.3 & 12.6 & BAT\\
100802A & 05:45:36 & $ 0^{\mathrm{h}} 09^{\mathrm{m}} 55^{\mathrm{s}}$ & $ 47^{\circ} 45' $ & H1L1$^\dagger$ & 36.4 & 16.3 & BAT\\
100804104 & 02:29:26 & $ 16^{\mathrm{h}} 35^{\mathrm{m}} 52^{\mathrm{s}}$ & $ 27^{\circ} 27' $ & H1L1 & 40.4 & 18.0 & GBM\\
100814A & 03:50:11 & $ 1^{\mathrm{h}} 29^{\mathrm{m}} 54^{\mathrm{s}}$ & $ -17^{\circ} 59' $ & H1L1V1$^\dagger$ & 17.3 & 7.3 & BAT\\
100814351 & 08:25:25 & $ 8^{\mathrm{h}} 11^{\mathrm{m}} 16^{\mathrm{s}}$ & $ 18^{\circ} 29' $ & L1V1$^*$ & 14.1 & 8.0 & GBM\\
100816009 & 00:12:41 & $ 6^{\mathrm{h}} 48^{\mathrm{m}} 28^{\mathrm{s}}$ & $ -26^{\circ} 40' $ & L1V1$^*$ & 6.6 & 3.5 & GBM\\
100819498 & 11:56:35 & $ 18^{\mathrm{h}} 38^{\mathrm{m}} 23^{\mathrm{s}}$ & $ -50^{\circ} 02' $ & H1L1V1 & 30.1 & 12.5 & GBM\\
100820373 & 08:56:58 & $ 17^{\mathrm{h}} 15^{\mathrm{m}} 09^{\mathrm{s}}$ & $ -18^{\circ} 31' $ & H1L1V1 & 18.2 & 7.2 & GBM\\
100823A & 17:25:35 & $ 1^{\mathrm{h}} 22^{\mathrm{m}} 49^{\mathrm{s}}$ & $ 5^{\circ} 51' $ & H1L1V1$^*$ & 8.7 & 4.0 & BAT\\
100825287 & 06:53:48 & $ 16^{\mathrm{h}} 53^{\mathrm{m}} 45^{\mathrm{s}}$ & $ -56^{\circ} 34' $ & H1L1V1 & 18.7 & 7.3 & GBM\\
100826A & 22:58:22 & $ 19^{\mathrm{h}} 05^{\mathrm{m}} 43^{\mathrm{s}}$ & $ -32^{\circ} 38' $ & L1V1 & 4.8 & 2.3 & GBM\\
100829876 & 21:02:08 & $ 6^{\mathrm{h}} 06^{\mathrm{m}} 52^{\mathrm{s}}$ & $ 29^{\circ} 43' $ & H1L1V1 & 12.3 & 4.7 & GBM\\
100904A & 01:33:43 & $ 11^{\mathrm{h}} 31^{\mathrm{m}} 37^{\mathrm{s}}$ & $ -16^{\circ} 11' $ & L1V1 & 13.1 & 7.3 & BAT\\
100905907 & 21:46:22 & $ 17^{\mathrm{h}} 30^{\mathrm{m}} 36^{\mathrm{s}}$ & $ 13^{\circ} 05' $ & H1L1V1 & 17.8 & 7.5 & GBM\\
100906A & 13:49:27 & $ 1^{\mathrm{h}} 54^{\mathrm{m}} 47^{\mathrm{s}}$ & $ 55^{\circ} 38' $ & H1L1$^*$$^\dagger$ & 30.5 & 12.2 & BAT\\
100916A & 18:41:12 & $ 10^{\mathrm{h}} 07^{\mathrm{m}} 50^{\mathrm{s}}$ & $ -59^{\circ} 23' $ & H1L1V1 & 8.3 & 3.6 & GBM\\
100917A & 05:03:25 & $ 19^{\mathrm{h}} 16^{\mathrm{m}} 59^{\mathrm{s}}$ & $ -17^{\circ} 07' $ & H1L1V1$^\dagger$ & 18.2 & 8.0 & BAT\\
100918863 & 20:42:18 & $ 20^{\mathrm{h}} 33^{\mathrm{m}} 38^{\mathrm{s}}$ & $ -45^{\circ} 58' $ & H1L1V1 & 19.4 & 7.8 & GBM\\
100919884 & 21:12:16 & $ 10^{\mathrm{h}} 52^{\mathrm{m}} 57^{\mathrm{s}}$ & $ 6^{\circ} 01' $ & H1L1V1 & 19.9 & 10.3 & GBM\\
100922625 & 14:59:43 & $ 23^{\mathrm{h}} 47^{\mathrm{m}} 55^{\mathrm{s}}$ & $ -25^{\circ} 11' $ & H1V1 & 9.2 & 4.7 & GBM\\
100926595 & 14:17:03 & $ 14^{\mathrm{h}} 50^{\mathrm{m}} 59^{\mathrm{s}}$ & $ -72^{\circ} 21' $ & H1L1 & 29.5 & 13.1 & GBM\\
100926694 & 16:39:54 & $ 2^{\mathrm{h}} 54^{\mathrm{m}} 19^{\mathrm{s}}$ & $ -11^{\circ} 06' $ & H1L1V1 & 12.2 & 5.1 & GBM\\
100929916 & 21:59:45 & $ 12^{\mathrm{h}} 12^{\mathrm{m}} 07^{\mathrm{s}}$ & $ -24^{\circ} 56' $ & H1V1 & 10.1 & 6.0 & GBM\\
101002279 & 06:41:26 & $ 21^{\mathrm{h}} 33^{\mathrm{m}} 23^{\mathrm{s}}$ & $ -27^{\circ} 28' $ & H1V1 & 9.5 & 4.6 & GBM\\
101003244 & 05:51:08 & $ 11^{\mathrm{h}} 43^{\mathrm{m}} 24^{\mathrm{s}}$ & $ 2^{\circ} 29' $ & H1L1V1 & 34.1 & 13.1 & GBM\\
101004426 & 10:13:49 & $ 15^{\mathrm{h}} 28^{\mathrm{m}} 52^{\mathrm{s}}$ & $ -43^{\circ} 59' $ & H1L1 & 46.8 & 17.5 & GBM\\
101010190 & 04:33:46 & $ 3^{\mathrm{h}} 08^{\mathrm{m}} 45^{\mathrm{s}}$ & $ 43^{\circ} 34' $ & L1V1 & 9.0 & 4.7 & GBM\\
101013412 & 09:52:42 & $ 19^{\mathrm{h}} 28^{\mathrm{m}} 19^{\mathrm{s}}$ & $ -49^{\circ} 38' $ & H1L1V1 & 35.8 & 15.0 & GBM\\
101015558 & 13:24:02 & $ 4^{\mathrm{h}} 52^{\mathrm{m}} 38^{\mathrm{s}}$ & $ 15^{\circ} 28' $ & H1L1 & 29.9 & 12.4 & GBM\\
101016243 & 05:50:16 & $ 8^{\mathrm{h}} 52^{\mathrm{m}} 09^{\mathrm{s}}$ & $ -4^{\circ} 37' $ & L1V1 & 9.4 & 4.7 & GBM\\
\end{longtable*}

\phantom{aaa}
\bibliographystyle{apj}

\begin{thebibliography}{144}
\expandafter\ifx\csname natexlab\endcsname\relax\def\natexlab#1{#1}\fi

\bibitem[{Aasi {et~al.}(2012)}]{VSR1-4DetChar}
Aasi, J., {et~al.} 2012, Class. Quantum Grav., 29, 155002

\bibitem[{Abadie {et~al.}(2010)}]{CBCrate}
Abadie, J., {et~al.} 2010, Class. Quant, Grav., 27, 173001

\bibitem[{Abadie {et~al.}(2012{\natexlab{a}})}]{2012arXiv1202.2788T}
---. 2012{\natexlab{a}}, Phys. Rev. D, 85, 122007

\bibitem[{Abadie {et~al.}(2012{\natexlab{b}})}]{grb051103}
---. 2012{\natexlab{b}}, Astrophys. J., 755, 2

\bibitem[{Abadie {et~al.}(2012{\natexlab{c}})}]{cbcLowMassS6}
---. 2012{\natexlab{c}}, Phys. Rev. D, 85, 082002

\bibitem[{Abbott {et~al.}(2004)}]{abbottnim04}
Abbott, B., {et~al.} 2004, Nucl. Inst. \& Meth. in Phys. Res., 517, 154

\bibitem[{Abbott {et~al.}(2005)}]{abbottgrb05}
---. 2005, Phys. Rev. D, 72, 042002

\bibitem[{Abbott {et~al.}(2008{\natexlab{a}})}]{grb070201_07}
---. 2008{\natexlab{a}}, Astrophys. J., 681, 1419

\bibitem[{Abbott {et~al.}(2008{\natexlab{b}})}]{burstGrbS234}
---. 2008{\natexlab{b}}, Phys. Rev. D, 77, 062004

\bibitem[{Abbott {et~al.}(2009{\natexlab{a}})}]{abbott-2007}
---. 2009{\natexlab{a}}, Rep. Prog. Phys., 72, 076901

\bibitem[{Abbott {et~al.}(2009{\natexlab{b}})}]{Abbott:2009tt}
Abbott, B.~P., {et~al.} 2009{\natexlab{b}}, Phys. Rev. D, 79, 122001

\bibitem[{Abbott {et~al.}(2010{\natexlab{a}})}]{burstGrbS5}
---. 2010{\natexlab{a}}, Astrophys. J., 715, 1438

\bibitem[{Abbott {et~al.}(2010{\natexlab{b}})}]{cbcGrbS5}
---. 2010{\natexlab{b}}, Astrophys. J, 715, 1453

\bibitem[{Accadia {et~al.}(2011)}]{VSR2calibration}
Accadia, T., {et~al.} 2011, Class. Quantum Grav., 28, 025005

\bibitem[{Accadia {et~al.}(2012)Accadia, Acernese, Alshourbagy, Amico,
  Antonucci, Aoudia, Arnaud, Arnault, Arun, Astone, Avino, Babusci, Ballardin,
  Barone, Barrand, Barsotti, Barsuglia, Basti, Bauer, Beauville, Bebronne,
  Bejger, Beker, Bellachia, Belletoile, Beney, Bernardini, Bigotta, Bilhaut,
  Birindelli, Bitossi, Bizouard, Blom, Boccara, Boget, Bondu, Bonelli, Bonnand,
  Boschi, Bosi, Bouedo, Bouhou, Bozzi, Bracci, Braccini, Bradaschia, Branchesi,
  Briant, Brillet, Brisson, Brocco, Bulik, Bulten, Buskulic, Buy, Cagnoli,
  Calamai, Calloni, Campagna, Canuel, Carbognani, Carbone, Cavalier, Cavalieri,
  Cecchi, Cella, Cesarini, Chassande-Mottin, Chatterji, Chiche, Chincarini,
  Chiummo, Christensen, Clapson, Cleva, Coccia, Cohadon, Colacino, Colas,
  Colla, Colombini, Conforto, Corsi, Cortese, Cottone, Coulon, Cuoco,
  D'Antonio, Daguin, Dari, Dattilo, David, Davier, Day, Debreczeni, Carolis,
  Dehamme, Fabbro, Pozzo, del Prete, Derome, Rosa, DeSalvo, Dialinas, Fiore,
  Lieto, Emilio, Virgilio, Dietz, Doets, Dominici, Dominjon, Drago, Drezen,
  Dujardin, Dulach, Eder, Eleuteri, Enard, Evans, Fabbroni, Fafone, Fang,
  Ferrante, Fidecaro, Fiori, Flaminio, Forest, Forte, Fournier, Fournier,
  Franc, Francois, Frasca, Frasconi, Freise, Gaddi, Galimberti, Gammaitoni,
  Ganau, Garnier, Garufi, Gáspár, Gemme, Genin, Gennai, Gennaro, Giacobone,
  Giazotto, Giordano, Giordano, Girard, Gouaty, Grado, Granata, Granata, Grave,
  Greverie, Groenstege, Guidi, Hamdani, Hayau, Hebri, Heidmann, Heitmann,
  Hello, Hemming, Hennes, Hermel, Heusse, Holloway, Huet, Iannarelli,
  Jaranowski, Jehanno, Journet, Karkar, Ketel, Voet, Kovalik, Kowalska,
  Kreckelbergh, Krolak, Lacotte, Lagrange, Penna, Laval, Marec, Leroy,
  Letendre, Li, Lieunard, Liguori, Lodygensky, Lopez, Lorenzini, Loriette,
  Losurdo, Loupias, Mackowski, Maiani, Majorana, Magazzù, Maksimovic,
  Malvezzi, Man, Mancini, Mansoux, Mantovani, Marchesoni, Marion, Marin,
  Marque, Martelli, Masserot, Massonnet, Matone, Matone, Mazzoni, Menzinger,
  Michel, Milano, Minenkov, Mitra, Mohan, Montorio, Morand, Moreau, Moreau,
  Morgado, Morgia, Mosca, Moscatelli, Mours, Mugnier, Mul, Naticchioni, Neri,
  Nocera, Pacaud, Pagliaroli, Pai, Palladino, Palomba, Paoletti, Paoletti,
  Paoli, Pardi, Parguez, Parisi, Pasqualetti, Passaquieti, Passuello,
  Perciballi, Perniola, Persichetti, Petit, Pichot, Piergiovanni, Pietka,
  Pignard, Pinard, Poggiani, Popolizio, Pradier, Prato, Prodi, Punturo, Puppo,
  Qipiani, Rabaste, Rabeling, Rácz, Raffaelli, Rapagnani, Rapisarda, Re,
  Reboux, Regimbau, Reita, Remilleux, Ricci, Ricciardi, Richard, Ripepe,
  Robinet, Rocchi, Rolland, Romano, Rosińska, Roudier, Ruggi, Russo, Salconi,
  Sannibale, Sassolas, Sentenac, Solimeno, Sottile, Sperandio, Stanga, Sturani,
  Swinkels, Tacca, Taddei, Taffarello, Tarallo, Tissot, Toncelli, Tonelli,
  Torre, Tournefier, Travasso, Tremola, Turri, Vajente, van~den Brand, Broeck,
  van~der Putten, Vasuth, Vavoulidis, Vedovato, Verkindt, Vetrano, Véziant,
  Viceré, Vinet, Vilalte, Vitale, Vocca, Ward, Was, Yamamoto, Yvert, Zendri,
  \& Zhang}]{1748-0221-7-03-P03012}
Accadia, T., Acernese, F., Alshourbagy, M., {et~al.} 2012, Journal of
  Instrumentation, 7, P03012

\bibitem[{Acernese {et~al.}(2007)Acernese, Amico, Alshourbagy, Antonucci,
  Aoudia, Astone, Avino, Babusci, Ballardin, Barone, Barsotti, Barsuglia,
  Bauer, Beauville, Bigotta, Birindelli, Bizouard, Boccara, Bondu, Bosi,
  Bradaschia, Braccini, van~den Brand, Brillet, Brisson, Buskulic, Calloni,
  Campagna, Carbognani, Cavalier, Cavalieri, Cella, Cesarini, Chassande-Mottin,
  Christensen, Corda, Corsi, Cottone, Clapson, Cleva, Coulon, Cuoco, Dari,
  Dattilo, Davier, del Prete, Rosa, Fiore, Virgilio, Dujardin, Eleuteri, Evans,
  Ferrante, Fidecaro, Fiori, Flaminio, Fournier, Frasca, Frasconi, Gammaitoni,
  Garufi, Genin, Gennai, Giazotto, Giordano, Giordano, Gouaty, Grosjean, Guidi,
  Hamdani, Hebri, Heitmann, Hello, Huet, Karkar, Kreckelbergh, Penna, Laval,
  Leroy, Letendre, Lopez, Lorenzini, Loriette, Losurdo, Mackowski, Majorana,
  Man, Mantovani, Marchesoni, Marion, Marque, Martelli, Masserot, Mazzoni,
  Milano, Menzinger, Moins, Moreau, Morgado, Mours, Nocera, Palomba, Paoletti,
  Pardi, Pasqualetti, Passaquieti, Passuello, Piergiovanni, Pinard, Poggiani,
  Punturo, Puppo, van~der Putten, Qipiani, Rapagnani, Reita, Remillieux, Ricci,
  Ricciardi, Ruggi, Russo, Solimeno, Spallicci, Tarallo, Tonelli, Toncelli,
  Tournefier, Travasso, Tremola, Vajente, Verkindt, Vetrano, Vicere, Vinet,
  Vocca, \& Yvert}]{Ac_etal:07}
Acernese, F., Amico, P., Alshourbagy, M., {et~al.} 2007, Class. Quantum Grav.,
  24, S671

\bibitem[{Acernese {et~al.}(2008)Acernese, Alshourbagy, Amico, Antonucci,
  Aoudia, Arun, Astone, Avino, Baggio, Ballardin, Barone, Barsotti, Barsuglia,
  Bauer, Bigotta, Birindelli, Bizouard, Boccara, Bondu, Bosi, Braccini,
  Bradaschia, Brillet, Brisson, Buskulic, Cagnoli, Calloni, Campagna,
  Carbognani, Cavalier, Cavalieri, Cella, Cesarini, Chassande-Mottin,
  Chatterji, Christensen, Cleva, Coccia, Corda, Corsi, Cottone, Coulon, Cuoco,
  D'Antonio, Dari, Dattilo, Davier, Rosa, Prete, Fiore, Lieto, Emilio,
  Virgilio, Evans, Fafone, Ferrante, Fidecaro, Fiori, Flaminio, Fournier,
  Frasca, Frasconi, Gammaitoni, Garufi, Genin, Gennai, Giazotto, Giordano,
  Granata, Greverie, Grosjean, Guidi, Hamdani, Hebri, Heitmann, Hello, Huet,
  Penna, Laval, Leroy, Letendre, Lopez, Lorenzini, Loriette, Losurdo,
  Mackowski, Majorana, Man, Mantovani, Marchesoni, Marion, Marque, Martelli,
  Masserot, Menzinger, Milano, Minenkov, Moins, Moreau, Morgado, Mosca, Mours,
  Neri, Nocera, Pagliaroli, Palomba, Paoletti, Pardi, Pasqualetti, Passaquieti,
  Passuello, Piergiovanni, Pinard, Poggiani, Punturo, Puppo, Rabaste,
  Rapagnani, Regimbau, Remillieux, Ricci, Ricciardi, Rocchi, Rolland, Romano,
  Ruggi, Russo, Sentenac, Solimeno, Swinkels, Terenzi, Toncelli, Tonelli,
  Tournefier, Travasso, Vajente, van~den Brand, van~der Putten, Verkindt,
  Vetrano, Vicere, Vinet, Vocca, \& Yvert}]{Ac_etal:08}
Acernese, F., Alshourbagy, M., Amico, P., {et~al.} 2008, Class. Quantum Grav.,
  25, 225001

\bibitem[{Acernese {et~al.}(2009)}]{aVirgo}
Acernese, F., {et~al.} 2009, {Virgo Technical Report VIR-0027A-09}

\bibitem[{Allen(2005)}]{Allen:2004gu}
Allen, B. 2005, Phys. Rev. D, 71, 062001

\bibitem[{Aloy {et~al.}(2000)Aloy, Muller, Ibanez, Marti, , \&
  MacFadyen}]{AlMuIbMaMa:00}
Aloy, M.~A., Muller, E., Ibanez, J.~M., {et~al.} 2000, Astrophys. J. Lett.,
  531, L119

\bibitem[{Barthelmy(2008)}]{Barthelmy08}
Barthelmy, S. 2008, Astronomische Nachrichten, 329, 340

\bibitem[{Barthelmy {et~al.}(2005)}]{Barthelmy:2005hs}
Barthelmy, S.~D., {et~al.} 2005, Space Science Reviews, 120

\bibitem[{Bartos {et~al.}(2011)}]{s6calibration}
Bartos, I., {et~al.} 2011, Frequency Domain Calibration Error Budget of LIGO
  Instruments in S6, Tech. rep., {LIGO-T1100071}

\bibitem[{Belczynski {et~al.}(2008)Belczynski, Taam, Rantsiou, \& van~der
  Sluys}]{Belczynski:2007xg}
Belczynski, K., Taam, R.~E., Rantsiou, E., \& van~der Sluys, M. 2008,
  Astrophys. J., 682, 474

\bibitem[{Berger {et~al.}(2005)}]{berger05}
Berger, E., {et~al.} 2005, Astrophys. J., 634, 501

\bibitem[{Biswas {et~al.}(2009)Biswas, Brady, Creighton, \&
  Fairhurst}]{Biswas:2007ni}
Biswas, R., Brady, P.~R., Creighton, J. D.~E., \& Fairhurst, S. 2009, Class.
  Quantum Grav., 26, 175009

\bibitem[{Blanchet(2006)}]{Blanchet:2006av}
Blanchet, L. 2006, Living Rev. Rel., 9, 3

\bibitem[{Blanchet {et~al.}(2004)Blanchet, Damour, Esposito-Far\`ese, \&
  Iyer}]{Blanchet:2004ek}
Blanchet, L., Damour, T., Esposito-Far\`ese, G., \& Iyer, B.~R. 2004, Phys.
  Rev. Lett., 93, 091101

\bibitem[{Blanchet {et~al.}(1995)Blanchet, Damour, Iyer, Will, \&
  Wiseman}]{Blanchet:1995ez}
Blanchet, L., Damour, T., Iyer, B.~R., Will, C.~M., \& Wiseman, A.~G. 1995,
  Phys. Rev. Lett., 74, 3515

\bibitem[{Bloom {et~al.}(2008)Bloom, Butler, \& Perley}]{Bloom:2008cn}
Bloom, J.~S., Butler, N.~R., \& Perley, D.~A. 2008, AIP Conf. Proc., 1000, 11

\bibitem[{Brady {et~al.}(2004)Brady, Creighton, \& Wiseman}]{Brady:2004gt}
Brady, P.~R., Creighton, J. D.~E., \& Wiseman, A.~G. 2004, Class. Quantum
  Grav., 21, S1775

\bibitem[{Burlon {et~al.}(2009)Burlon, Ghirlanda, Ghisellini, Greiner, \&
  Celotti}]{Burlon09}
Burlon, D., Ghirlanda, G., Ghisellini, G., Greiner, J., \& Celotti, A. 2009,
  Astron. Astrophys., 505, 569

\bibitem[{Burlon {et~al.}(2008)}]{Burlon08}
Burlon, D., {et~al.} 2008, Astrophys. J. Lett., 685, L19

\bibitem[{Burrows {et~al.}(2006)}]{Burrows:2006ar}
Burrows, D.~N., {et~al.} 2006, Astrophys. J., 653, 468

\bibitem[{{Chapman} {et~al.}(2009){Chapman}, {Priddey}, \&
  {Tanvir}}]{2009MNRAS.395.1515C}
{Chapman}, R., {Priddey}, R.~S., \& {Tanvir}, N.~R. 2009, Mon. Not. R. Astron.
  Soc., 395, 1515

\bibitem[{{Chapman} {et~al.}(2007){Chapman}, {Tanvir}, {Priddey}, \&
  {Levan}}]{2007MNRAS.382L..21C}
{Chapman}, R., {Tanvir}, N.~R., {Priddey}, R.~S., \& {Levan}, A.~J. 2007, Mon.
  Not. R. Astron. Soc., 382, L21

\bibitem[{{Chernoff} \& {Finn}(1993)}]{1993ApJ...411L...5C}
{Chernoff}, D.~F., \& {Finn}, L.~S. 1993, \apjl, 411, L5

\bibitem[{Cokelaer(2007)}]{hexabank}
Cokelaer, T. 2007, Phys.~Rev.~D, 76, 102004

\bibitem[{Connaughton(2011)}]{GBMsysErr_GCN}
Connaughton, V. 2011, {GCN circular 11574}

\bibitem[{Corsi \& Mészáros(2009)}]{Corsi09}
Corsi, A., \& Mészáros, P. 2009, Astrophys. J., 702, 1171

\bibitem[{Cutler \& Flanagan(1994)}]{Cutler:1994ys}
Cutler, C., \& Flanagan, E.~E. 1994, Phys. Rev., D49, 2658

\bibitem[{Dalal {et~al.}(2006)Dalal, Holz, Hughes, \& Jain}]{Dalal:2006qt}
Dalal, N., Holz, D.~E., Hughes, S.~A., \& Jain, B. 2006, Phys. Rev. D, 74,
  063006

\bibitem[{Davies {et~al.}(2002)Davies, King, Rosswog, \& Wynn}]{Davies02}
Davies, M.~B., King, A., Rosswog, S., \& Wynn, G. 2002, Astrophys. J. Lett.,
  579, L63

\bibitem[{Davies {et~al.}(2005)Davies, Levan, \& King}]{Davies:2004pu}
Davies, M.~B., Levan, A.~J., \& King, A.~R. 2005, Mon. Not. R. Astron. Soc.,
  356, 54

\bibitem[{Dietz(2011)}]{Dietz:2010eh}
Dietz, A. 2011, Astron. Astrophys., 529, A97

\bibitem[{Dragoljub(1993)}]{Dragoljub93}
Dragoljub, M. 1993, Phys. Rev. D, 48, 4738

\bibitem[{Duez(2010)}]{0264-9381-27-11-114002}
Duez, M.~D. 2010, Classical and Quantum Gravity, 27, 114002

\bibitem[{{Duncan} \& {Thompson}(1992)}]{duncan92}
{Duncan}, R.~C., \& {Thompson}, C. 1992, Astrophys. J., 392, L9

\bibitem[{Eichler {et~al.}(1989)Eichler, Livio, Piran, \& Schramm}]{schramm89}
Eichler, D., Livio, M., Piran, T., \& Schramm, D.~N. 1989, Nature, 340, 126

\bibitem[{Etienne {et~al.}(2008)}]{Etienne:2007jg}
Etienne, Z.~B., {et~al.} 2008, Phys. Rev. D, 77, 084002

\bibitem[{Faber {et~al.}(2006)Faber, Baumgarte, Shapiro, Taniguchi, \&
  Rasio}]{Faber:2006tx}
Faber, J.~A., Baumgarte, T.~W., Shapiro, S.~L., Taniguchi, K., \& Rasio, F.~A.
  2006, AIP Conf. Proc., 861, 622

\bibitem[{Feroci {et~al.}(2007)}]{superAGILE07}
Feroci, A., {et~al.} 2007, Nucl. Instrum. Meth. A, 581, 728

\bibitem[{Ferrari {et~al.}(2010)Ferrari, Gualtieri, \&
  Pannarale}]{Ferrari:2009bw}
Ferrari, V., Gualtieri, L., \& Pannarale, F. 2010, Phys.Rev., D81, 064026

\bibitem[{Finn \& Chernoff(1993)}]{Finn:1992xs}
Finn, L.~S., \& Chernoff, D.~F. 1993, Phys. Rev., D47, 2198

\bibitem[{Flanagan \& Hinderer(2008)}]{Flanagan:2007ix}
Flanagan, E.~E., \& Hinderer, T. 2008, Phys. Rev., D77, 021502

\bibitem[{Foucart {et~al.}(2011)Foucart, Duez, Kidder, \&
  Teukolsky}]{Foucart11}
Foucart, F., Duez, M.~D., Kidder, L.~E., \& Teukolsky, S.~A. 2011, Phys. Rev.
  D, 83, 024005

\bibitem[{Fox {et~al.}(2005)}]{fox05}
Fox, D.~B., {et~al.} 2005, Nature, 437, 845

\bibitem[{Frederiks {et~al.}(2007)Frederiks, Palshin, Aptekar, Golenetskii,
  Cline, \& Mazets}]{Frederiks07}
Frederiks, D., Palshin, V., Aptekar, R., {et~al.} 2007, Astronomy Letters, 33,
  19

\bibitem[{Fryer {et~al.}(2002)Fryer, Holz, \& Hughes}]{Fryer02}
Fryer, C.~L., Holz, D.~E., \& Hughes, S.~A. 2002, Astrophys.J., 565, 430

\bibitem[{Gal-Yam(2006)}]{Gal-Yam06}
Gal-Yam, A. 2006, Astrophys. J., 639, 331

\bibitem[{Galama {et~al.}(1998)}]{galama98}
Galama, T.~J., {et~al.} 1998, Nature, 395, 670

\bibitem[{Gao {et~al.}(2010)Gao, Lu, \& Zhang}]{Gao:2010yh}
Gao, H., Lu, Y., \& Zhang, S.~N. 2010, Astrophys. J., 717, 268

\bibitem[{Gehrels {et~al.}(2009)Gehrels, Ramirez-Ruiz, \& Fox}]{Gehrels09}
Gehrels, N., Ramirez-Ruiz, E., \& Fox, D.~B. 2009, Annu. Rev. Astron. Astr.,
  47, 567

\bibitem[{{Gehrels} {et~al.}(2004){Gehrels}, {Chincarini}, {Giommi}, {Mason},
  {Nousek}, {Wells}, {White}, {Barthelmy}, {Burrows}, {Cominsky}, {Hurley},
  {Marshall}, {M{\'e}sz{\'a}ros}, {Roming}, {Angelini}, {Barbier}, {Belloni},
  {Campana}, {Caraveo}, {Chester}, {Citterio}, {Cline}, {Cropper}, {Cummings},
  {Dean}, {Feigelson}, {Fenimore}, {Frail}, {Fruchter}, {Garmire}, {Gendreau},
  {Ghisellini}, {Greiner}, {Hill}, {Hunsberger}, {Krimm}, {Kulkarni}, {Kumar},
  {Lebrun}, {Lloyd-Ronning}, {Markwardt}, {Mattson}, {Mushotzky}, {Norris},
  {Osborne}, {Paczynski}, {Palmer}, {Park}, {Parsons}, {Paul}, {Rees},
  {Reynolds}, {Rhoads}, {Sasseen}, {Schaefer}, {Short}, {Smale}, {Smith},
  {Stella}, {Tagliaferri}, {Takahashi}, {Tashiro}, {Townsley}, {Tueller},
  {Turner}, {Vietri}, {Voges}, {Ward}, {Willingale}, {Zerbi}, \&
  {Zhang}}]{swift04}
{Gehrels}, N., {Chincarini}, G., {Giommi}, P., {et~al.} 2004, Astrophys. J.,
  611, 1005

\bibitem[{Gehrels {et~al.}(2006)}]{Gehrels:2006tk}
Gehrels, N., {et~al.} 2006, Nature, 444, 1044

\bibitem[{Grote {et~al.}(2008)}]{geo08}
Grote, H., {et~al.} 2008, Class. Quantum Grav., 25, 114043

\bibitem[{Grupe {et~al.}(2006)}]{Grupe:2006uc}
Grupe, D., {et~al.} 2006, Astrophys. J., 653, 462

\bibitem[{{Guetta} \& {Piran}(2005)}]{GuPi:05}
{Guetta}, D., \& {Piran}, T. 2005, Astron. \& Astrophys., 435, 421

\bibitem[{Hanna(2008)}]{Hanna:2008}
Hanna, C. 2008, PhD thesis, Louisiana State University

\bibitem[{Harry {et~al.}(2010)}]{aLIGO}
Harry, G.~M., {et~al.} 2010, Class. Quantum Grav., 27, 084006

\bibitem[{Harry \& Fairhurst(2011)}]{Harry:2010fr}
Harry, I.~W., \& Fairhurst, S. 2011, Phys. Rev., D83, 084002

\bibitem[{Helmstrom(1968)}]{helmstrom-1968}
Helmstrom, C.~W. 1968, Statistical Theory of Signal Detection, 2nd edition
  (Pergamon Press, London)

\bibitem[{Hessels {et~al.}(2006)Hessels, Ransom, Stairs, Freire, Kaspi,
  {et~al.}}]{Hessels:2006ze}
Hessels, J.~W., Ransom, S.~M., Stairs, I.~H., {et~al.} 2006, Science, 311, 1901

\bibitem[{{Hinderer} {et~al.}(2010){Hinderer}, {Lackey}, {Lang}, \&
  {Read}}]{2010PhRvD..81l3016H}
{Hinderer}, T., {Lackey}, B.~D., {Lang}, R.~N., \& {Read}, J.~S. 2010, \prd,
  81, 123016

\bibitem[{Hjorth \& Bloom(2011)}]{2011arXiv1104.2274H}
Hjorth, J., \& Bloom, J.~S. 2011, {Gamma-Ray Bursts} (Cambridge University
  Press), arXiv:1104.2274

\bibitem[{Horvath {et~al.}(2010)}]{Horvath:2010um}
Horvath, I., {et~al.} 2010, Astrophys. J., 713, 552

\bibitem[{Hurley {et~al.}(2009)}]{ipn2009}
Hurley, K., {et~al.} 2009, in American Institute of Physics Conference Series,
  Vol. 1133, GAMMA-RAY BURST: Sixth Huntsville Symposium, ed. C.~Meegan,
  C.~Kouveliotou, \& N.~Gehrels, 55--57

\bibitem[{{Hurley} {et~al.}(2010){Hurley}, {Rowlinson}, {Bellm}, {Perley},
  {Mitrofanov}, {Golovin}, {Kozyrev}, {Litvak}, {Sanin}, {Boynton}, {Fellows},
  {Harshmann}, {Ohno}, {Yamaoka}, {Nakagawa}, {Smith}, {Cline}, {Tanvir},
  {O'Brien}, {Wiersema}, {Rol}, {Levan}, {Rhoads}, {Fruchter}, {Bersier},
  {Kavelaars}, {Gehrels}, {Krimm}, {Palmer}, {Duncan}, {Wigger}, {Hajdas},
  {Atteia}, {Ricker}, {Vanderspek}, {Rau}, \& {von
  Kienlin}}]{2010MNRAS.403..342H}
{Hurley}, K., {Rowlinson}, A., {Bellm}, E., {et~al.} 2010, \mnras, 403, 342

\bibitem[{Iwamoto {et~al.}(1998)}]{iwamoto98}
Iwamoto, K., {et~al.} 1998, Nature, 395, 672

\bibitem[{Jakobsson {et~al.}(2006)}]{Jakobsson06}
Jakobsson, P., {et~al.} 2006, Astron. Astrophys., 447, 897

\bibitem[{Jakobsson {et~al.}(2012)}]{Jakobsson12}
---. 2012, Astrophys. J., in press

\bibitem[{Kiziltan {et~al.}(2010)Kiziltan, Kottas, \&
  Thorsett}]{Kiziltan:2010ct}
Kiziltan, B., Kottas, A., \& Thorsett, S.~E. 2010, arXiv:1011.4291
  [astro-ph.GA]

\bibitem[{Kobayashi \& Mészáros(2003{\natexlab{a}})}]{Kobayashi03}
Kobayashi, S., \& Mészáros, P. 2003{\natexlab{a}}, Astrophys. J., 589, 861

\bibitem[{Kobayashi \& Mészáros(2003{\natexlab{b}})}]{Kobayashi03-1}
---. 2003{\natexlab{b}}, Astrophys. J. Lett., 585, L89

\bibitem[{Kochanek \& Piran(1993)}]{kochanek93}
Kochanek, C.~S., \& Piran, T. 1993, Astrophys. J., 417, L17

\bibitem[{Komatsu {et~al.}(2011)}]{Komatsu10}
Komatsu, E., {et~al.} 2011, Astrophys. J. Suppl. S., 192, 18

\bibitem[{{Koshut} {et~al.}(1995){Koshut}, {Kouveliotou}, {Paciesas}, {van
  Paradijs}, {Pendleton}, {Briggs}, {Fishman}, \&
  {Meegan}}]{1995ApJ...452..145K}
{Koshut}, T.~M., {Kouveliotou}, C., {Paciesas}, W.~S., {et~al.} 1995, \apj,
  452, 145

\bibitem[{{Kouveliotou} {et~al.}(1993){Kouveliotou}, {Meegan}, {Fishman},
  {Bhat}, {Briggs}, {Koshut}, {Paciesas}, \& {Pendleton}}]{ck93}
{Kouveliotou}, C., {Meegan}, C.~A., {Fishman}, G.~J., {et~al.} 1993, Astrophys.
  J., 413, L101

\bibitem[{Kulkarni {et~al.}(1998)}]{kulkarni98}
Kulkarni, S.~R., {et~al.} 1998, Nature, 395, 663

\bibitem[{Lackey {et~al.}(2012)Lackey, Kyutoku, Shibata, Brady, \&
  Friedman}]{2011arXiv1109.3402L}
Lackey, B.~D., Kyutoku, K., Shibata, M., Brady, P.~R., \& Friedman, J.~L. 2012,
  Phys. Rev. D, 85, 044061

\bibitem[{Lazzati(2005)}]{Lazzati05}
Lazzati, D. 2005, Mon. Not. R. Astron. Soc., 357, 722

\bibitem[{Lazzati {et~al.}(2009)Lazzati, Morsony, , \& Begelman}]{LaMoBe:09}
Lazzati, D., Morsony, B.~J., , \& Begelman, M.~C. 2009, Astrophys. J. Lett.,
  700, L47

\bibitem[{Le \& Dermer(2007)}]{Le07}
Le, T., \& Dermer, C.~D. 2007, Astrophys. J., 661, 394

\bibitem[{Leonor {et~al.}(2009)}]{Leonor09}
Leonor, I., {et~al.} 2009, Class. Quantum. Grav, 26, 204017

\bibitem[{{Liang} {et~al.}(2007){Liang}, {Zhang}, {Virgili}, \&
  {Dai}}]{2007ApJ...662.1111L}
{Liang}, E., {Zhang}, B., {Virgili}, F., \& {Dai}, Z.~G. 2007, Astrophys. J.,
  662, 1111

\bibitem[{MacFadyen {et~al.}(2001)MacFadyen, Woosley, \& Heger}]{MacFadyen01}
MacFadyen, A.~I., Woosley, S.~E., \& Heger, A. 2001, Astrophys. J., 550, 410

\bibitem[{Mandel \& O'Shaughnessy(2010)}]{Mandel:2009nx}
Mandel, I., \& O'Shaughnessy, R. 2010, Class. Quant. Grav., 27, 114007

\bibitem[{Matsuoka {et~al.}(2009)}]{MAXI09}
Matsuoka, M., {et~al.} 2009, Publ. Astron. Soc. Japan, 61, 999

\bibitem[{{Mazets} {et~al.}(2008){Mazets}, {Aptekar}, {Cline}, {Frederiks},
  {Goldsten}, {Golenetskii}, {Hurley}, {von Kienlin}, \& {Pal'shin}}]{mazets07}
{Mazets}, E.~P., {Aptekar}, R.~L., {Cline}, T.~L., {et~al.} 2008, Astrophys.
  J., 680, 545

\bibitem[{Meegan {et~al.}(2009)}]{GBM09}
Meegan, C., {et~al.} 2009, Astrophys. J., 702, 791

\bibitem[{M{\'{e}}sz{\'{a}}ros(2006)}]{Meszaros:2006rc}
M{\'{e}}sz{\'{a}}ros, P. 2006, Rept. Prog. Phys., 69, 2259

\bibitem[{Metzger \& Berger(2012)}]{Metzger11}
Metzger, B.~D., \& Berger, E. 2012, Astrophys. J., 746, 48

\bibitem[{{Modjaz}(2011)}]{2011AN....332..434M}
{Modjaz}, M. 2011, Astronomische Nachrichten, 332, 434

\bibitem[{Nakar(2007)}]{nakar-2007}
Nakar, E. 2007, Physics Reports, 442, 166

\bibitem[{{Nakar} {et~al.}(2006){Nakar}, {Gal-Yam}, \& {Fox}}]{NaGaFo:06}
{Nakar}, E., {Gal-Yam}, A., \& {Fox}, D.~B. 2006, Astrophys. J., 650, 281

\bibitem[{Narayan {et~al.}(1992)Narayan, Paczynski, \& Piran}]{Narayan:1992iy}
Narayan, R., Paczynski, B., \& Piran, T. 1992, Astrophys. J., 395, L83

\bibitem[{Nissanke {et~al.}(2010)Nissanke, Holz, Hughes, Dalal, \&
  Sievers}]{Nissanke:2009kt}
Nissanke, S., Holz, D.~E., Hughes, S.~A., Dalal, N., \& Sievers, J.~L. 2010,
  Astrophys. J., 725, 496

\bibitem[{Norris \& Bonnell(2006)}]{Norris06}
Norris, J.~P., \& Bonnell, J.~T. 2006, Astrophys. J., 643, 266

\bibitem[{Ott(2009)}]{Ott09-1}
Ott, C.~D. 2009, Class. Quant. Grav., 26, 063001

\bibitem[{Ott {et~al.}(2006)Ott, Burrows, Dessart, \& Livne}]{Ott06}
Ott, C.~D., Burrows, A., Dessart, L., \& Livne, E. 2006, Phys. Rev. Lett., 96,
  201102

\bibitem[{Ozel {et~al.}(2012)Ozel, Psaltis, Narayan, \&
  Villarreal}]{Ozel:2012ax}
Ozel, F., Psaltis, D., Narayan, R., \& Villarreal, A.~S. 2012, arXiv:1201.1006

\bibitem[{Paciesas {et~al.}(2012)}]{Paciesas12}
Paciesas, W.~S., {et~al.} 2012, Astrophys. J. Suppl. Ser., 199, 18

\bibitem[{{Palmer} {et~al.}(2005){Palmer}, {Barthelmy}, {Gehrels}, {Kippen},
  {Cayton}, {Kouveliotou}, {Eichler}, {Wijers}, {Woods}, {Granot}, {Lyubarsky},
  {Ramirez-Ruiz}, {Barbier}, {Chester}, {Cummings}, {Fenimore}, {Finger},
  {Gaensler}, {Hullinger}, {Krimm}, {Markwardt}, {Nousek}, {Parsons}, {Patel},
  {Sakamoto}, {Sato}, {Suzuki}, \& {Tueller}}]{palmer05}
{Palmer}, D.~M., {Barthelmy}, S., {Gehrels}, N., {et~al.} 2005, Nature, 434,
  1107

\bibitem[{{Pannarale} {et~al.}(2011){Pannarale}, {Rezzolla}, {Ohme}, \&
  {Read}}]{2011PhRvD..84j4017P}
{Pannarale}, F., {Rezzolla}, L., {Ohme}, F., \& {Read}, J.~S. 2011, \prd, 84,
  104017

\bibitem[{Piro \& Pfahl(2007)}]{Piro07}
Piro, A.~L., \& Pfahl, E. 2007, Astrophys. J., 658, 1173

\bibitem[{Poisson \& Will(1995)}]{Poisson:1995ef}
Poisson, E., \& Will, C.~M. 1995, Phys.~Rev.~D, 52, 848

\bibitem[{Predoi \& Hurley(2012)}]{Predoi:2011aa}
Predoi, V., \& Hurley, K. 2012, J. Phys. Conf. Ser., 363, 012034

\bibitem[{Racusin {et~al.}(2009)}]{Racusin09}
Racusin, J.~L., {et~al.} 2009, Astrophys. J., 698, 43

\bibitem[{Rantsiou {et~al.}(2008)Rantsiou, Kobayashi, Laguna, \&
  Rasio}]{Rantsiou:2007ct}
Rantsiou, E., Kobayashi, S., Laguna, P., \& Rasio, F.~A. 2008, Astrophys. J.,
  680, 1326

\bibitem[{Read {et~al.}(2009)}]{Read:2009yp}
Read, J.~S., {et~al.} 2009, Phys. Rev., D79, 124033

\bibitem[{Rezzolla {et~al.}(2011)}]{Rezzolla11}
Rezzolla, L., {et~al.} 2011, Astrophys. J. Lett., 732, 1

\bibitem[{Romero {et~al.}(2010)Romero, Reynoso, \& Christiansen}]{Romero10}
Romero, G.~E., Reynoso, M.~M., \& Christiansen, H.~R. 2010, Astron. Astrophys.,
  524, A4

\bibitem[{Rosswog(2006)}]{Rosswog:2006ue}
Rosswog, S. 2006, Rev. Mex. Astron. Astrofis., 27, 57, arXiv:astro-ph/0612572

\bibitem[{Schutz(1986)}]{Schutz:1986gp}
Schutz, B.~F. 1986, Nature, 323, 310

\bibitem[{Shibata {et~al.}(2003)Shibata, Shigeyuki, \& Yoshiharu}]{Shibata03}
Shibata, M., Shigeyuki, K., \& Yoshiharu, E. 2003, Mon. Not. R. Astron. Soc.,
  343, 619

\bibitem[{Shibata \& Taniguchi(2008)}]{Shibata:2007zm}
Shibata, M., \& Taniguchi, K. 2008, Phys. Rev., D77, 084015

\bibitem[{Shibata \& Taniguchi(2011)}]{lrr-2011-6}
---. 2011, Living Reviews in Relativity, 14

\bibitem[{Soderberg {et~al.}(2006)}]{NatureSoderberg2006}
Soderberg, A.~M., {et~al.} 2006, Nature, 442, 1014

\bibitem[{Sutton {et~al.}(2010)}]{Sutton:2009gi}
Sutton, P.~J., {et~al.} 2010, New J. Phys., 12, 053034

\bibitem[{{Tanvir} {et~al.}(2005){Tanvir}, {Chapman}, {Levan}, \&
  {Priddey}}]{2005Natur.438..991T}
{Tanvir}, N.~R., {Chapman}, R., {Levan}, A.~J., \& {Priddey}, R.~S. 2005, \nat,
  438, 991

\bibitem[{Thorne(1987)}]{thorne.k:1987}
Thorne, K.~S. 1987, in Three hundred years of gravitation, ed. S.~W. Hawking \&
  W.~Israel (Cambridge: Cambridge University Press), 330--458

\bibitem[{{Vallisneri}(2000)}]{2000PhRvL..84.3519V}
{Vallisneri}, M. 2000, Physical Review Letters, 84, 3519

\bibitem[{Vedrenne \& Atteia(2009)}]{Vedrenne:ch5}
Vedrenne, G., \& Atteia, J.-L. 2009, Gamma-Ray Bursts (Springer), 219--252

\bibitem[{Virgili {et~al.}(2009)Virgili, Liang, \& Zhang}]{Virgili07}
Virgili, F.~J., Liang, E.-W., \& Zhang, B. 2009, Mon. Not. R. Astron. Soc.,
  392, 91

\bibitem[{Wang \& Meszaros(2007)}]{WaMe:07}
Wang, X.-Y., \& Meszaros, P. 2007, Astrophys. J., 670, 1247

\bibitem[{Watson {et~al.}(2007)Watson, Fynbo, Thone, \&
  Sollerman}]{Watson:2007nk}
Watson, D., Fynbo, J. P.~U., Thone, C.~C., \& Sollerman, J. 2007, Phil. Trans.
  Roy. Soc. Lond., A365, 1269

\bibitem[{Will(2005)}]{Will:2005va}
Will, C.~M. 2005, Living Rev. Rel., 9, 3

\bibitem[{Winkler {et~al.}(2003)}]{integral03}
Winkler, C., {et~al.} 2003, Astron. \& Astrophys., 411, L1

\bibitem[{Woosley(2012)}]{Woosley11}
Woosley, S.~E. 2012, Gamma-ray Bursts (Cambridge University Press),
  arXiv:1105.4193

\bibitem[{Wąs(2011)}]{thesisWas}
Wąs, M. 2011, PhD thesis, Laboratoire de l'Accélerateur Linéaire, {LAL
  11-119}

\bibitem[{Wąs {et~al.}(2012)Wąs, Sutton, Jones, \& Leonor}]{Was12}
Wąs, M., Sutton, P.~J., Jones, G., \& Leonor, I. 2012, Phys. Rev. D, 86,
  022003

\bibitem[{Zhang {et~al.}(2007)}]{Zhang:2006mb}
Zhang, B., {et~al.} 2007, Astrophys. J., 655, L25

\bibitem[{Zhang {et~al.}(2009)}]{Zhang:2009uf}
---. 2009, Astroph. J., 703, 1696

\bibitem[{Zhang {et~al.}(2003)Zhang, Woosley, , \& MacFadyen}]{ZhWoMa:03}
Zhang, W., Woosley, S.~E., , \& MacFadyen, A.~I. 2003, Astrophys. J., 586, 356

\end{thebibliography}

\end{document}